\documentclass{aa}
\usepackage{graphicx}
\usepackage{rotating}
\usepackage{amsmath}
\usepackage{txfonts}
\usepackage{natbib}
\bibpunct{(}{)}{;}{a}{}{,}

\begin{document}

\title{High energy afterglows and flares from gamma-ray burst by inverse Compton emission}
\author{A.~Galli\inst{1}$^{, }$\inst{2}$^{, }$\inst{3}
\and L.~Piro\inst{1} } 

\offprints{A. Galli: alessandra.galli@iasf-roma.inaf.it} 

\institute{IASF-Roma/INAF, via fosso del cavaliere 100, 00133 Roma, Italy 
	\and
          Universit\'a degli Studi di Roma "La Sapienza", Piazzale A. Moro 5, 00185, Roma, Italy
	\and
          INFN - Sezione di Trieste, Padriciano 99, 34021 Trieste, Italy}

\date{Received ....; accepted ....}


\abstract{We perform a detailed study of inverse Compton (IC) emission for a fireball undergoing external shock (ES) in either a uniform or a wind-like interstellar medium, and assess the relative importance of IC and synchrotron emissions. We determine the primary model parameters driving the IC to synchrotron emission ratio in the case of a short duration central engine. We then investigate the case of ES by a long duration central engine, or delayed external shock (DES), a model that can account for some of the flares observed in  gamma-ray bursts (GRB) X-ray light curves at late times. We present model predictions, in particular in terms of GeV vs X-ray behavior, and compare them with other models proposed to explain the origin of flares. We find that if most of the emission occurs when the fireball is in the fast cooling regime, then a substantial GeV emission is expected both for a short (standard ES) and a long (DES) duration central engine activity. In particular, in the context of standard ES we are able to account for the delayed emission observed in GRB940217. In the case of DES, we find that IC scattering of X-ray flare photons can produce high energy flares in the GeV band, which can be detected by the Large Area Telescope aboard GLAST. The detectability of high energy flares improves with the burst kinetic energy $E$: about 30 \% of \emph{Swift} GRBs showing flares in their X-ray light curve have sufficiently large kinetic energy so that the expected high flares can be detected by GLAST. One important prediction of the DES model is the simultaneity between low and high energy flares. To test this simultaneity, the peak energies of both flares need to fall below or within the observational bands. We predict that X-ray flares with peak energy of $\sim$ 10 eV produce high energy flares with peak energy of around 100 MeV-GeV. Observations by \emph{Swift} and GLAST then, can test the predicted simultaneity, thus helping to discriminate between different models.

\keywords{radiation mechanism: non-thermal --- gamma rays: burst}
}

\authorrunning{A. Galli \& L. Piro}

\titlerunning{GRB high energy emission}

\maketitle


\section{Introduction}

One of the key areas that would clarify the nature of gamma-ray burst (GRB) progenitors is to understand the mechanism(s) acting during the GRB prompt-to-afterglow transition phase. This phase takes place from hundreds to thousands of seconds after the burst, and is characterized by a variety of temporal and spectral behavior due to the contribution of both prompt and afterglow emissions. Particularly interesting is the appearance of X-ray flares. The comprehension of this phenomenon can give important information on the physics of the central engine.
\emph {Swift} observations showed that X-ray flares are present in a large fraction of GRBs (about one half of the \emph{Swift} GRB sample) and occur both in long and short bursts \citep{obrien06}. X-ray flares exhibit a different behavior in terms of their spectral properties: most of them present hard-to-soft spectral evolution resembling that of the prompt emission \citep{burrows05,butler07}, while others seem to exhibit a softer and not evolving spectrum consistent with that of the subsequent afterglow emission (e.g. \citet{piro05,galli06}). Several models have been proposed in the literature to explain the origin of GRB flaring activity. The most important difference between these models is that some of them do not require a long duration central engine activity (group 1), while other require a long duration and/or re-activation of the central engine (group 2). Models which belong to group 1 are late internal shocks (LIS) from a short duration central engine \citep{zhang05,wu05}, refreshed shocks \citep{rees98,kumar00refresh,guetta06}, two components jet \citep{meszaros01,lipunov01}, patchy jet \citep{kumar00}, forward shock (FS)-reverse shock (RS) \citep{fan05}, external shock (ES) with a clumpy medium \citep{dermer00,dermer07} and delayed magnetic dissipation in strongly magnetized ejecta caused by external shock \citep{giannios06}. 

In the scenario of LIS from a short duration central engine, in addition to the shells producing the prompt-gamma ray emission through internal shocks, a tail of slow (small Lorentz factor) shells is emitted by the central engine. These slower shells can collide (internal dissipation) at a later time thus producing (X-ray) flares.  In this case a late prompt emission is produced, while there is no external shock emission because the medium has been previously cleaned up by the faster shells \citep{lazzati06}. In this context, in order to explain a delayed X-ray emission appearing hundreds to thousands of seconds after the burst, a small difference in the Lorentz factor $\Gamma$ of the colliding shells would be necessary, i.e. $\Delta \Gamma << \Gamma$. However a small contrast between shells cannot explain the fast temporal variability typically observed in X-ray flares (see \citet{krimm07} for details). Refreshed shocks occurs when the central engine releases its energy with a variety of Lorentz factors, thus the faster part of the outflow decelerates earlier and the slower part can catch up with it at later times (when it has decelerated due to the interaction with the external medium), injecting energy into the blast wave and producing a flare in the light curve. In the framework of two-components jet the flare is produced by the deceleration of a moderately relativistic jet component as it interacts with the external medium. Patchy jets are characterized by large energy fluctuations in the angular direction and can be imagined as multi-component jets, thus they can produce several bumps in GRB light curves. However, these bumps are typically too shallow to explain the fast rise and decay observed in X-ray flares \citep{zhang05}. \citet{fan05} have shown that a FS-RS scenario can also explain the appearance of bumps in GRBs light curves; for appropriate model parameters, synchrotron RS emission can dominate in the X-ray band, producing a signature in the light curve. In the framework of ES, the interaction of the fireball with clouds of sufficiently small radii can produce high variable GRB light curves and a delayed flaring activity \citep{dermer00,dermer07}. Finally, \citet{giannios06} showed that magnetic dissipation during external shocks can produce fast-evolving and energetic flares if the dissipation regions have appropriate dimensions.

Within group 2 models, X-ray flares can be produced by LIS \citep{burrows05,wu05} or  ES \citep{piro05,galli06,panaitescu07}. The prolonged central engine activity could be due to a long-lived accretion of blobs of material on the central black hole \citep{king05,perna06}. In the context of LIS, X-ray flares are considered prompt emission, that lasts for hundreds to thousands of seconds. We note that, in such a case, late internal shocks have to be tuned to produce X-ray emission rather than gamma-ray emission. In the case of ES by a long duration central engine the flare represents the onset of afterglow emission. This condition is verified if the central engine releases a significant fraction of the energy at late times, and when its duration is longer than the deceleration time (the so called thick shell case). In this case the onset of afterglow emission is delayed by several hundreds of seconds; we refer to this model as delayed external shock (DES). Recently, \citet{panaitescu07} has shown that X-ray flares could also be produced through the up-scattering of the forward shock emission by a more relativistic and lepton-enhanced shell emitted at the time of the flare appearance.

X-ray flares appear mostly during the first phases of afterglow emission; consequently, independently on their origin, the delayed X-ray flare photons are expected to interact with the afterglow electrons by inverse Compton (IC) giving rise to delayed high energy counterparts. In this work we study flares in the context of DES and assume an external medium with a smooth density profile (i.e a uniform interstellar medium (ISM) and wind-like medium). We decide to study flares in the context of ES, motivated by the spectral similarities observed in some bursts between X-ray flares and afterglow emission, which can be straightforwardly explained in this scenario. This scenario, depending on the spectral index of the electron population, can also account for those flares showing a spectral evolution of the order of 0.5-1.0, if the typical emission frequency is passing through the observational band. \emph{Swift} observations showed that several bursts present multiple flares in their X-ray light curve, e.g. \object{GRB 050730} \citep{pandey06}. In these cases, we can explain only one flare because only one flare can represent the onset of afterglow emission, and the other flares have to be ascribed to the interaction of the fireball with a clumpy medium (in the framework of ES), or to any other of the models of group 1 and/or 2.

 In the DES scenario the emission mechanism responsible for X-ray flares is synchrotron, and the flare photons can be IC up-scattered, thus producing flares in the GeV band. In this case, X-ray and high energy flares are produced by the same emitting region and electrons population, thus one expect that - depending on the energy - the two flares have similar temporal profiles and no significant delay. These predictions can be tested during the GLAST - \emph{Swift} era and play an important role in order to discriminate between several models proposed to explain the origin of flares. Indeed, high energy flares can be produced also in the context of LIS models \citep{wang06,fan06,fan07}. \citet{wang06} have proposed two possible mechanisms that are able to produce flares. In the first one, X-ray flares are produced through synchrotron emission in LIS, and the flare photons are self-IC scattered by the same electrons, producing the flare. In this case, the X-ray and the high energy flare come from the same region, thus they are expected to have a good temporal correlation as in the ES case. However, in the context of LIS, low Lorentz factors are required to produce long timescale X-ray flares \citep{falcone05,wang06,wu05}, and consequently less energetic flares are expected in the GeV band in comparison with the ES model. In the second mechanism, X-ray flares are produced by first-order IC emission from LIS, and these photons have a second order IC scatter with the afterglow electrons. X-ray flare photons need time to reach and scatter with the afterglow electrons. During this time the beam spreads out, and this causes a delayed and longer high energy flare (for  a detailed analysis see \citet{fan07}). In addition, in this case we still expect high energy flares of lower intensity with respect to those expected by the ES.

The paper is organized as follows; in Sect. \ref{contour} we study IC versus synchrotron emission in the so called standard case of a thin shell fireball. We first follow an analytical approach that can be applied during the fireball deceleration phase (Sect. \ref{comparison}). We evaluate the effect of pair production on the emitted radiation, and study the detectability of the IC component above the GeV band by the Large Area Telescope (LAT) aboard GLAST. We also develop a numerical model that permit the study of the fireball at times preceding the deceleration phase (Sect. \ref{thin}). In Sect. \ref{940217} we show that an ES by a thin shell fireball with synchrotron plus IC emission is able to model the delayed high energy emission observed in \object{GRB 940217}. We study the properties of X-ray flares in the context of the DES (thick shell fireball), and determine the burst properties improving the detection of high energy flares, see Sect. \ref{predizionihighen}. In Sect. \ref{thick} we first provide analytical formulas that allow one to predict, under reasonable assumptions, whether an X-ray flare with given observed properties can produce a detectable high energy flare. We then present a numerical study of X-ray and high energy flares by a DES (thick shell fireball), and apply this model to the individual cases of \object{XRF 011030} and \object{GRB 011121} (Sect. \ref{varicasi}). We present our conclusions in Sect. \ref{conclusionigen}.

\section{Inverse Compton emission for a standard thin shell fireball in the framework of external shock.}
\label{contour}

The emission mechanism explaining GRB prompt and afterglow radiation is synchrotron emission from relativistic electrons accelerated in internal and external shocks, respectively. We assume that the electrons are accelerated to a single relativistic power law distribution of index $p$, i.e. $N(\gamma)$ $\propto \gamma^{-p}$, with electron minimum energy $\gamma_i = (m_p / m_e) \epsilon_e (\Gamma -1)$, where $\Gamma$ is the fireball Lorentz factor and $\epsilon_e$ is the fraction of the fireball energy going into relativistic electrons \citep{panaitescu00}. Photons produced by synchrotron emission can interact with the electrons by IC (or self synchrotron Compton: SSC), giving rise to a high-energy emission component.

The synchrotron emission spectrum is described by four power law segments separated by the three characteristic frequencies, the absorption frequency $\nu_a$ , the injection frequency $\nu_i$ and the cooling frequency $\nu_c$, during both the slow and fast cooling regimes. The IC component has a spectrum similar to that of the synchrotron component with the characteristic frequencies boosted by a factor $\sim \gamma^2$: $\nu_{a,IC}= 2 \gamma_i^2 \nu_a$, $\nu_{i,IC}= 2 \gamma_i^2 \nu_i$ and $\nu_{c,IC}= 2 \gamma_c^2 \nu_{c}$ in the Thompson regime. \citet{esin01} have shown that at frequencies greater than $\nu_{i,IC}$ the flux has to be increased by a logarithmic term with respect to the power law approximation. For simplicity, in this work we follow the prescriptions of \citet{panaitescu00} and adopt for the IC spectrum a power law approximation, which is also a more conservative choice.

\subsection{A comparative study of Inverse Compton versus synchrotron emission}
\label{comparison}

In this section we study the relative importance of IC and synchrotron emission as a function of the parameters of the Fireball Model, i.e. the energy injected into the fireball $E_{53}$, the external medium density $n$ or the parameter $A_*$ in the case of a wind-like medium, and the efficiencies $\epsilon_e$ and $\epsilon_B$. We initially study synchrotron and IC emission at time scales larger than the fireball deceleration time when analytical formulas can be used (for a detailed numerical study see Sect. \ref{thin}).

We also require that IC emission occurs in the Thompson regime, i.e that the energy of target photons $\gamma_i (h \nu)$ is smaller than the electron rest energy $m_e c^2$. This condition translates into $\gamma_i < \gamma_{i,max}= \Gamma m_e c^2$/$(h \nu_x)$ \citep{wang06}, where $\gamma_i$ is the minimum post-shock electrons Lorentz factor, $\gamma_{i,max}$ is the maximum electrons Lorentz factor and $h \nu_x$ is the peak energy of the synchrotron emission. Using the prescriptions of \citet{panaitescu00} we find that the condition that IC emission occurs in the Thompson regime translates into:

\begin{equation}
\label{eelim}
\epsilon_{e,lim} < 0.28 \biggl( \frac{h \nu_x}{1 keV} \biggr)^{-1}
\end{equation}

either for a fireball expanding in a uniform or a wind-like interstellar medium. The relative importance of IC and synchrotron emission is given by the Compton parameter $Y$, defined as the ratio of IC and synchrotron luminosities $L_{syn}$ and $L_{IC}$, see Eq. (3.1) of \citet{esin01}: 

\begin{equation}
\label{eq:lumic}
Y=L_{IC}/L_{syn} .
\end{equation}

We compute $Y$ according to Eq. 35 of \citet{panaitescu00}, correcting the $(\epsilon_e/\epsilon_B)$ ratio for the radiative efficiency $\eta$ (the efficiency $\eta$ is the fraction of energy radiated via synchrotron and IC):

\begin{equation}
Y = \frac{1}{2}\biggl[ \biggl( \frac{5-s}{2(3-s)} \frac{\epsilon_e}{\epsilon_B} \eta + 1 \biggr)^{1/2}-1 \biggr]
\end{equation}

with $s$=0(2) for an ISM (wind), $\eta=1$ during the fast cooling regime and $\eta=(\nu_c/\nu_i)^{(2-p)/2}$ during the slow cooling regime. $Y$ is proportional to $(\eta \epsilon_e/\epsilon_B)^{1/2}$ if $\eta \epsilon_e/\epsilon_B >> 1$ and to $(\eta \epsilon_e/\epsilon_B)$ if $\eta \epsilon_e/\epsilon_B < 1$ \citep{esin01}.

\begin{figure}[!htb]
\centering
\includegraphics[width=8.5cm,height=8.5cm]{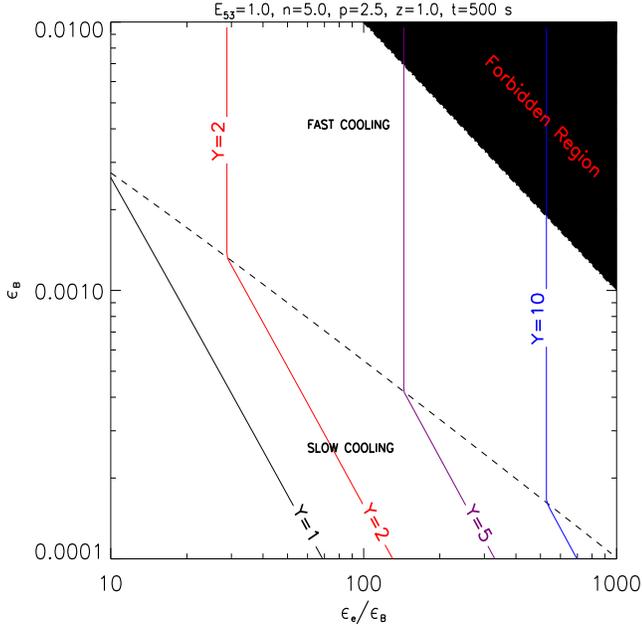}
\caption{Regions in the $(\epsilon_e/\epsilon_B)$, $\epsilon_B$ parameters space where the IC component dominates over the synchrotron component at $t=500$ s for a fireball expanding in an ISM. The other model parameters are fixed to $E_{53}=1.0$, $n=5.0$, and $p=2.5$. The redshift is $z=1$. The black solid line corresponds to $Y=1$, the red solid line corresponds to $Y=2$, the purple solid line to $Y=5$ and the blue solid line to $Y=10$. The dashed line separates the region of the parameters space where the fireball is in the fast cooling regime from that where it is in the slow cooling regime. }
\label{contourism}
\end{figure}

Clearly, the model parameters that mainly determine the relative importance of IC and synchrotron emission are $\epsilon_e$ and $\epsilon_B$. We first study the importance of IC emission as a function of $\epsilon_e$ and $\epsilon_B$ in the case of a fireball expanding in an ISM, and set the remaining model parameters to $E_{53}=1.0$ , $n=5.0$, $p=2.5$, and fix the redshift to $z=1$. In Fig. \ref{contourism} we plot Y as a function of $(\epsilon_e/\epsilon_B)$ and $\epsilon_B$ at $t=500$ s. With the parameters above and an initial fireball Lorentz factor $\Gamma_0$=100, the fireball deceleration time is $t_{dec} \sim $ 250 s \citep{panaitescu00}, i.e at $t=500$ s the fireball is at the beginning of the deceleration phase.

The red solid line in Fig. \ref{contourism} corresponds to $Y=2$, the purple solid line to $Y=5$ and the blue solid line to $Y=10$, while the black solid line is the line of IC and synchrotron equal luminosity.

The black region in Fig. \ref{contourism} is forbidden because it implies $\epsilon_e >$1. We have also checked that the Thompson regime condition (that we express through Eq. \ref{eelim}) is verified for any choice of parameters. The dashed line in Fig. \ref{contourism} separates the region of parameters space where the fireball is in the fast cooling regime from that where the fireball is in the slow cooling regime, and corresponds to values of $\epsilon_B$ and $(\epsilon_e/\epsilon_B)$ that satisfy the following equation:

\begin{equation}\label{eq:transizioneism}
\epsilon_B=7.96 \times 10^{-2}(E_{53}n)^{-1/4}(1+Y)^{-1/2}(\epsilon_e/\epsilon_B)^{-1/2}T_d^{1/4}(1+z)^{-1/4}
\end{equation}

where $T_d$ is the time expressed in unity of 1 day. This equation tells us that for a given set of model parameters the size of the fast cooling region decreases with time, and that at a given time the higher the values of $E_{53}$ and  $n$ are, the larger the size of the fast cooling region is. For a given $\epsilon_B$ and $(\epsilon_e/\epsilon_B)$ the relative importance of IC emission with respect to synchrotron emission is greater during the fast cooling regime, thus IC emission is favored by high values of $E_{53}$ and a medium with large density n.
The solid lines in Fig. \ref{contourism} show that during the fast cooling regime, the relative importance of IC and synchrotron emission depends only on the $(\epsilon_e/\epsilon_B)$ ratio and increases with $(\epsilon_e/\epsilon_B)$ for any values of $\epsilon_B$. During the slow cooling regime $\eta < 1.0$, and to achieve the same value of $Y$ greater values of $(\epsilon_e/\epsilon_B)$ are required.

\begin{figure}[!htb]
\centering
\includegraphics[width=8.5cm,height=8.5cm]{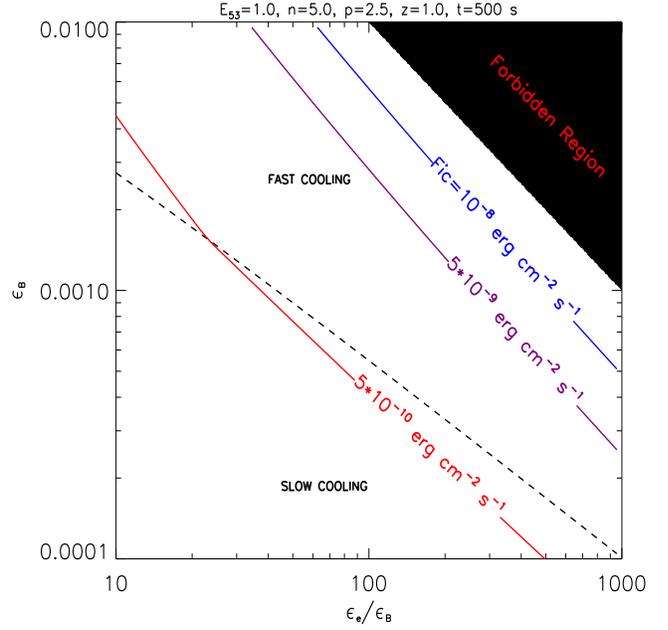}
\caption{IC flux in the $(\epsilon_e/\epsilon_B)$, $\epsilon_B$ parameters space at $t=500$ s for a fireball expanding in a ISM. The other model parameters are the same as Fig. \ref{contourism}. The dashed line separates the region of the parameters space where the fireball is in the fast cooling regime from that where it is in the slow cooling regime. The red solid curve refers to an IC flux $F_{IC}= 5 \times 10^{-10}~erg~cm^{-2} s^{-1}$, the purple solid line refers to $F_{IC}= 5 \times 10^{-9}~erg~cm^{-2} s^{-1}$, and the blue solid line refers to $F_{IC}= 10^{-8}~erg~cm^{-2} s^{-1}$.}
\label{contourismlum}
\end{figure}

We also study how the $\nu F_{\nu}$ IC flux varies in the $(\epsilon_e/\epsilon_B)$, $\epsilon_B$ parameters space (see Fig. \ref{contourismlum}). We compute IC flux using Eq. \ref{eq:lumic}. In this equation, synchrotron and IC luminosity should be calculated by integration over the total spectrum, but one can evaluate them and give order of magnitude estimates computing synchrotron and IC flux at the peak of the spectrum, i.e $L_{syn} \simeq \nu_{max}F_{\nu}(\nu_{max})$ and $L_{IC} \simeq \nu_{IC,max}F_{\nu_{IC}}(\nu_{IC,max})$, with $\nu_{max}=\nu_i$ during the fast cooling regime and $\nu_{max}=\nu_c$ during the slow cooling regime.
For a fireball expanding in an ISM we find:
\begin{equation}
\begin{split}
\label{lumicfastvalue}
(\nu F_{\nu})_{IC}& = 2.4 \times 10^{-10} D_{l,28}^{-2}E_{53}\epsilon_B \bigg( \frac{\epsilon_e}{\epsilon_B} \bigg) (1+Y)^{-1}Y\\
      & \quad T_d^{-1}(1+z)~ erg~cm^{-2}s^{-1}
\end{split}
\end{equation}  

during the fast cooling regime and

\begin{equation}
\begin{split}
\label{lumicslowvalue}
(\nu F_{\nu})_{IC}& = 3 \times 10^{-9} D_{l,28}^{-2}E_{53}^{13/8}n^{1/4}(1+Y)^{-1/2}Y\epsilon_B^{2} \bigg( \frac{\epsilon_e}{\epsilon_B} \bigg)^{3/2}\\
      & \quad T_d^{-5/4}(1+z)^{5/4}~ erg~cm^{-2}s^{-1}
\end{split}
\end{equation}  

during the slow cooling regime for $p=2.5$. During the fast cooling regime the IC flux increases with $\epsilon_e$. It does not depend on either $\epsilon_B$ nor on the density $n$ of the external medium. This implies that the estimates presented in Fig. \ref{contourismlum} are not strongly affected by the uncertainties regarding the external medium density n. During the slow cooling regime the radiative efficiency $\eta$ is proportional to $n^{(p-2)/2} (1+Y)^{(p-2)}$ and $Y \sim n^{(p-2)/2(4-p)}$, thus from Eq. \ref{lumicslowvalue} we find $(\nu F_{\nu})_{IC} \sim n^{(p-2)/(4-p)}$, i.e IC flux increases with n.

In the case of a fireball expanding in a wind-like medium we find similar results, i.e the Compton parameter Y and the IC peak flux have similar behaviors as a function of the model parameters and similar values to those found in the ISM case. The fast-to-slow cooling transition occurs for:

\begin{equation}
\label{transitionwind} 
\epsilon_B=6.6 \times 10^{-2} A_*^{-1/2} (1+Y)^{-1/2} \bigg( \frac{\epsilon_e}{\epsilon_B} \bigg)^{-1/2} T_d^{1/2} (1+z)^{-1/2}
\end{equation}

This equation tells us that for a fixed time, the higher the wind density parameter $A_*$ is, the larger the region of the parameters space where the fireball is in the fast cooling regime is, i.e. IC emission is also favored by large densities when the fireball expands in a medium with a wind density profile. Contrary to the ISM case, the extension of the fast cooling region does not depend on the fireball energy $E_{53}$. Using the same approach followed for the ISM case, we find that for a fireball expanding in a wind medium density profile the IC flux is:

\begin{equation}
\begin{split}
\label{lumicfastwind}
(\nu F_{\nu})_{IC}& = 4.7 \times 10^{-10} D_{l,28}^{-2} E_{53} \epsilon_B \biggl(\frac{\epsilon_e}{\epsilon_B}\biggr) (1+Y)^{-1}Y\\
      & \quad T_d^{-1} (1+z)~ erg~cm^{-2}s^{-1}
\end{split}
\end{equation}  

during the fast cooling regime and

\begin{equation}
\begin{split}
\label{lumicslowwind}
(\nu F_{\nu})_{IC}& = 7.2 \times 10^{-9} D_{l,28}^{-2} E_{53} A_*^{1/2} \epsilon_B^2 \biggl(\frac{\epsilon_e}{\epsilon_B}\biggr)^{3/2} (1+Y)^{-1/2}Y T_d^{-3/2}\\
      & \quad (1+z)^{3/2}~ erg~cm^{-2}s^{-1}\\
\end{split}
\end{equation}  

during the slow cooling (for $p=2.5$). For a fireball expanding in a wind-like medium during the fast cooling regime, $(\nu F_{\nu})_{IC}$ increases with $\epsilon_e$, while it does not depend on $\epsilon_B$ and the wind parameter $A_*$, as for the ISM case. During the slow cooling regime the radiative efficiency $\eta$ is proportional to $A_*^{(p-2)}(1+Y)^{(p-2)}$ and $Y \sim A_*^{(p-2)/(4-p)}$, thus from Eq. \ref{lumicslowwind} we find $(\nu F_{\nu})_{IC} \sim A_*^{2(p-2)/(4-p)}$, i.e. during the slow cooling regime $(\nu F_{\nu})_{IC}$ increases with $A_*$.

\subsection{Pair attenuation estimates}
\label{pairattenuation}

Here we estimate the attenuation of IC emission by pair production due to $\gamma \gamma$ interactions. This process could be very important because it causes internal absorption which determines a cutoff in the spectrum of the emitted radiation at high energies. Given a photon of energy $E_\gamma$, the threshold energy $E_{th}$ for the production of an electron-positron pair is:

\begin{equation}\label{pairsoglia}
E_{th} = \frac{2 (m_ec^2)^2 \Gamma^2}{E_{\gamma} (1 -cos \theta)}
\end{equation}

where $\theta$ is the impact angle between the two photons in the shell frame. Taking into account the prescriptions of \citet{panaitescu00}, for a fireball expanding in an ISM we find that the threshold frequency $\nu_{th}$ where pair production becomes important is:

\begin{equation}\label{freqsoglia}
\nu_{th}=2.5 \times 10^{15} E_{53}^{1/4} n^{-1/4} T_d^{-3/4} (1+z)^{-5/4} \biggl( \frac{E_\gamma}{1TeV} \biggr)^{-1}~Hz
\end{equation}

The optical depth to pair production for a photon of energy $E_{\gamma}$ is (see Eq. (2) of \citet{wang04}:

\begin{equation}\label{profondita2}
\tau_{\gamma \gamma}=\biggl(\frac{11}{180}\biggr)\frac{\sigma_T D_{lum}^2 F_{\nu}(\nu_{th})}{4 \Gamma^4 c^2 h \beta t}
\end{equation}

where $(11/180)\sigma_T$ is an analytical approximation for the $\gamma \gamma$ cross section \citep{svensson87}, $D_{lum}$ is the source distance and $\beta$ is the spectral index of the radiation emitted by the source. During the first phases of afterglow emission, i.e. around $\sim 500$s, for typical GRB model parameters $\nu_c < \nu_i < \nu_{th}$ thus $F_{\nu}(\nu_{th})= F_{\nu,max}(\nu_{th}/\nu_i)^{-p/2}(\nu_c/\nu_i)^{1/2}$ and $\beta=p/2$. When $\tau_{\gamma \gamma}=1$ the pair attenuation cannot be neglected, and the spectrum is limited by a cutoff energy $E_{cut}$:

\begin{equation}
\label{cutoffism}
E_{cut} \simeq 0.5 E_{53}^{-1/4} n^{-\frac{(p+4)}{4p}}\epsilon_{e,-1}^{-\frac{2(p-1)}{p}}\epsilon_{B,-2}^{-\frac{(p-2)}{2p}} (1+Y)^{2/p}T_d^{-\frac{8-3p}{4p}}(1+z)^{\frac{(8-7p)}{4p}}~TeV
\end{equation}

Similarly, for a fireball expanding in a wind-like medium we find the cutoff energy to be:

\begin{equation}
\label{cutoffwind}
E_{cut} \simeq 0.9 E_{53}^{1/p} A_*^{-\frac{(p+4)}{2p}}\epsilon_{e,-1}^{-\frac{2(p-1)}{p}}\epsilon_{B,-2}^{-\frac{(p-2)}{2p}} (1+Y)^{2/p} T_d^{-\frac{3-p}{p}}(1+z)^{-\frac{(2p-1)}{p}}~TeV
\end{equation}

For typical model parameters $E_{cut} \sim$TeV, around $500$ s, both for a fireball expanding in an ISM and in a wind-like medium; consequently our model predictions (which do not account for the pair attenuation) should not in general be affected by internal absorption up to energies of the order of TeV. We note also that for typical model parameters around 1 day the cutoff energy is still above the LAT energy band.


\subsection{Detectability of the Inverse Compton component in the GeV band with the GLAST Large Area Telescope}
\label{latsensitivity}

Now we compare the IC flux predicted in the context of the ES by a thin shell fireball with the GLAST Large Area Telescope (LAT) sensitivity during both early and late afterglow emission. We estimate the LAT sensitivity adopting the same criterion of \citet{zhang04}, that a 5$\sigma$ detection is made when at least 5 photons are collected if the instrument is source dominated. We evaluate the LAT sensitivity between 100 MeV and 200 GeV, the spectral range where the instrument point spread function (PSF) and effective area $A_{eff}$ are well characterized. We integrate at each energy E the burst spectrum in an energy range $\Delta E=E/2$, and also take into account the time when the LAT becomes background dominated.  When the instrument is source dominated the LAT sensitivity $F_{th}$ decreases linearly with time according to the following law:

\begin{equation}
\label{soglialat}
F_{th}(E)=\frac{5}{A_{eff}(E) t}~cm^{-2} s^{-1} 
\end{equation}

where $E$ is the photon energy and $t$ is the integration time.  The most important contribution to the background comes from the observed high latitude diffuse flux, $\sim 1.5 \times 10^{-5} cm^{-2} s^{-1} sr^{-1}$ above 100 MeV, see http://www-glast.slac.stanford.edu/software/IS/glast\_lat\_performance.htm. When the instrument is no longer source dominated its sensitivity starts to decrease with $t^{1/2}$. The sensitivity curve is displayed with a black dot-dashed line in Figs. \ref{spettro200s_ismthickfast_vp10ev} and \ref{spettro200s_ismthickfast_vp1kev} for a short integration time (500 s), and in Figs. \ref{spettro10000_ismthinfast} and \ref{spettro10000_windthinfast} for a long integration time (10000 s). We also evaluate the LAT sensitivity threshold as a function of the integration time at 1 GeV, the observational energy where we display the predicted light curves, see Sects. \ref{thin} and \ref{thick}. At 1 GeV the on-axis LAT effective area is $A_{eff} \simeq 0.85~m^2$ and the PSF is of the order of 1$^{\circ}$, thus the LAT sensitivity is $\sim 1.9 \times 10^{-9}~erg~cm^{-2}s^{-1}$ for an integration time $t=500$ s, and is $\sim 9.4 \times 10^{-11}~erg~cm^{-2}s^{-1}$ for an integration time $t=10000$ s, and the LAT is source dominated up to an integration time of $\sim$ 61500 s.

\begin{figure}[!htb]
\centering
\includegraphics[width=8.5cm,height=8.5cm]{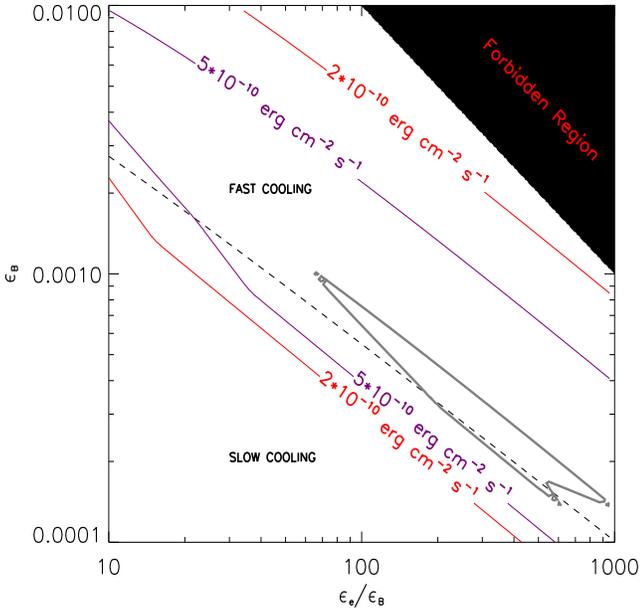}
\caption{$\nu F_{\nu}$ of the IC at 1 GeV in the $(\epsilon_e/\epsilon_B)$, $\epsilon_B$ parameters space at $t=500$ s for a fireball expanding in an ISM. The other model parameters are the same as in Fig. \ref{contourism}. The dashed line separates the region of the parameters space where the fireball is in the fast cooling regime from that where it is in the slow cooling regime. The grey solid line represents the LAT sensitivity for an integration time of 500 s. The purple solid line refers to an IC flux $(\nu F_{\nu})_{IC}=5 \times 10^{-10}~erg~ cm^{-2}s^{-1}$, and the red solid line to $(\nu F_{\nu})_{IC}=2 \times 10^{-10}~erg~ cm^{-2}s^{-1}$. }
\label{contourismlum1gev}
\end{figure}

\begin{figure}[!htb]
\centering
\includegraphics[width=8.5cm,height=8.5cm]{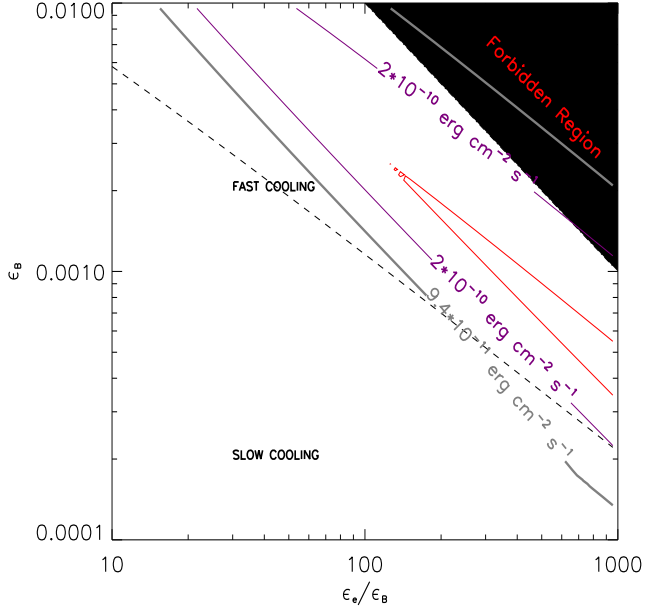}
\caption{$\nu F_{\nu}$ of the IC at 1 GeV in the $(\epsilon_e/\epsilon_B)$, $\epsilon_B$ parameters space at $t=10000$ s for a fireball expanding in an ISM. The other model parameters are the same as in Fig. \ref{contourism}. The dashed line separates the region of the parameters space where the fireball is in the fast cooling regime from that where it is in the slow cooling regime. The grey solid line represents the LAT sensitivity for an integration time of 10000 s. The red solid line refers to an IC flux $(\nu F_{\nu})_{IC}=2 \times 10^{-10}~erg~ cm^{-2}s^{-1}$, and the purple solid line to $(\nu F_{\nu})_{IC}=5 \times 10^{-10}~erg~ cm^{-2}s^{-1}$. }
\label{contourismlum1gev10000sec}
\end{figure}

We compute the IC flux for a burst located at redshift z=1 at the observational energy of 1 GeV (following the prescriptions of \citet{panaitescu00}). In Fig. \ref{contourismlum1gev} we compare our predictions with the 1 GeV LAT sensitivity (grey solid line) for an integration time of 500 s. It shows that LAT will be able to detect IC emission only during the fast cooling regime, and for a very narrow region of the $(\epsilon_e/\epsilon_B)$, $\epsilon_B$ parameters space. However, if the integration time is sufficiently long (but the instrument is still source dominated) LAT will be able to detect IC emission for a significantly larger region of the parameters space, see Fig. \ref{contourismlum1gev10000sec}. We find similar results for the detectability of the IC component associated with a fireball expanding in a medium with a wind density profile; also in this case the detectability of the IC component increases with the integration time if the LAT is source dominated. 
                
We can make similar considerations for the detectability of the IC component by the gamma-ray imaging detector (GRID) aboard AGILE (Astro-rivelatore Gamma a Immagini LEggero; launched April 23, 2007), which is sensitive in the range 30 MeV-50 GeV (see \citet{tavani} for a recent review). To make a comparison between the GRID and the LAT sensitivity it is important to know the time when the GRID becomes background dominated. At early times the GRID is source dominated and we can evaluate its sensitivity using Eq. \ref{soglialat} as for the LAT.  At late times the GRID is background dominated and its sensitivity decreases as $t^{-1/2}$.  We take the GRID sensitivity at 50 hours from \citet{tavani} and we find that it lies above the extrapolation of eq. \ref{soglialat}. This implies that at 50 hours the GRID is background dominated. The transition from one regime to the other takes place around 2000 s. The GRID on axis effective area at 1 GeV is $\sim 550~ cm^{-2}$ which implies that its sensitivity is $\sim 15$ times lower than that of the LAT, i.e $\sim 2.9 \times 10^{-8}~erg~ cm^{-2}s^{-1}$ for an integration time of 500 s. For longer integration times, when the GRID becomes background dominated, its sensitivity goes as $F_{th}=5.5 \times 10^{-7}t^{-1/2}$, which gives a flux threshold of $\sim 5.5 \times 10^{-9}~erg~ cm^{-2}s^{-1}$ for an integration time of 10000 s.
A comparison of GRID sensitivity with Figs. \ref{contourismlum1gev} and \ref{contourismlum1gev10000sec} shows that it will not be able to detect ES IC emission for a burst located at z=1 for both short and longer integration times. This is consistent with the findings of \citet{gou07}.

\subsection{Inverse Compton spectra and light curves}
\label{thin}
Here we would like to asses in detail the maximum contribution of IC emission and how it varies with time in the framework of standard ES (i.e thin shell fireballs). Analytical equations are valid only at times larger then the fireball deceleration time and permit us to model only the fading portion of the burst light curve. Now we adopt a numerical approach which allow us to describe also the rising part of the light curve (see \citet{galli06} for details).

We start with the case of a thin shell fireball expanding in an ISM, which remains in the fast cooling regime during a large time interval of afterglow emission, e.g. up to thousands of seconds after the burst, because this is the condition that maximizes the IC component. In Sect. \ref{comparison} we have shown that during the fast cooling regime the relative importance of IC and synchrotron emission is determined only by the $(\epsilon_e/\epsilon_B)$ ratio, and the IC flux increases linearly with $\epsilon_e$ while substantially does not depend on $\epsilon_B$. We thus assume a large (but reasonable) value of $\epsilon_e$ that also satisfies the request that IC emission occurs in the Thompson regime (see Eq. \ref{eelim}), and vary the value of $\epsilon_B$. We choose the other model parameters consistently with the requests of a thin shell and fast cooling regime. Equation \ref{eq:transizioneism} tells us that the time the fireball remains in the fast cooling regime increases with $E_{53}$ and $n$, and Eq. \ref{lumicfastvalue} tells us that IC emission increases with $E_{53}$; this motivates us to assume large values of fireball energy and external medium density such as $E_{53}$=1 and $n=5$. In an ISM the thin shell condition reads \citep{piro05}:

\begin{equation}
\label{thinism}
T_2 \Bigg( \frac{n}{E_{53}} \Bigg)^{1/3}\Gamma_{0,2}^{8/3} < 6.2
\end{equation}

where $T_2$ is the prompt emission duration in unity of 100 s and $\Gamma_{0,2}$ is the initial fireball Lorentz factor in unity of 100. We fix $\Gamma_0=120$, which ensures the thin shell condition is satisfied for a prompt emission duration as long as $\sim$ 200 s. We note that the value of $\Gamma_0$ affects only the first phases of ES emission, while during the deceleration phase, neither synchrotron nor IC emission depend on $\Gamma_0$. Finally, we fix the redshift $z$ and the spectral electron population index $p$ to canonical values as $z$=1 and $p$=2.5 \citep{catalogosax}.

\begin{figure}[!htb]
\centering
\includegraphics[scale=0.35,angle=-90]{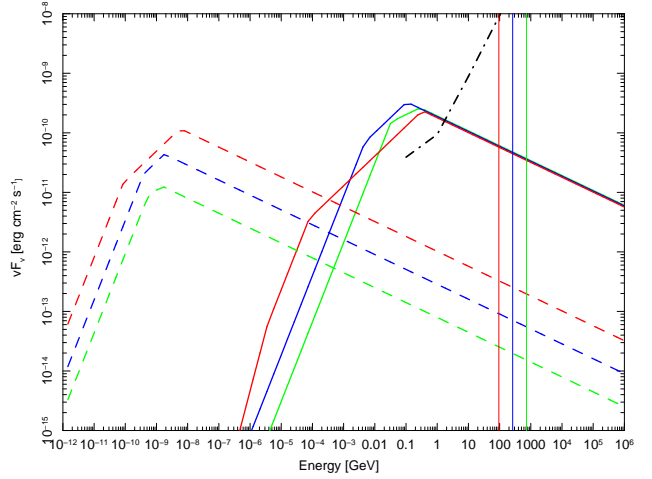}
\caption{Synchrotron (dashed lines) and IC (solid lines) spectra for a thin shell fireball expanding in an ISM at {\bf $t \sim 10000$ s}. The model parameters are $\Gamma_0$=120, $E_{53}=1.0$, $n=5$ and $\epsilon_e=0.2$. Green, blue and red spectra are obtained respectively for $\epsilon_B=0.0001$, $\epsilon_B=0.001$, and $\epsilon_B=0.01$. The thin solid lines indicate the cutoff energy due to pair production for the three IC spectra. The black dot-dashed line is the 5 $\sigma$ LAT sensitivity for an integration time of $\sim$ 10000 s.}
\label{spettro10000_ismthinfast}
\end{figure}

\begin{figure}[!htb]
\centering
\includegraphics[scale=0.35,angle=-90]{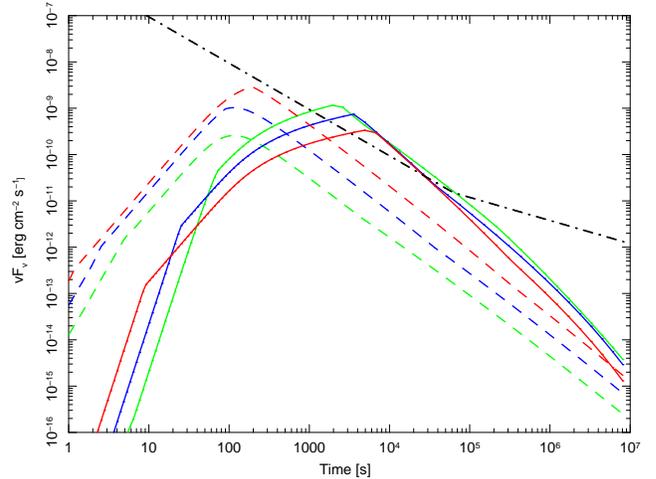}
\caption{Synchrotron (dashed line) and IC (solid line) light curves for a thin shell fireball expanding in an ISM. The model parameters are the same as in Fig. \ref{spettro10000_ismthinfast}. Synchrotron light curves are computed at $10^{18}$ Hz and IC light curves are computed at $2.4 \times 10^{23}$Hz. The green curves correspond to $\epsilon_B=0.0001$, the blue curves correspond to $\epsilon_B=0.001$, and the red curves correspond to $\epsilon_B=0.01$. For these choices of model parameters at $\sim 500$ s the fireball is in the fast cooling regime. The black dot-dashed line shows the {\bf 5 $\sigma$} LAT threshold at 1 GeV as a function of the integration time.}
\label{thin_fast_Eb}
\end{figure}

We present the effect of ($\epsilon_e/\epsilon_B$) on synchrotron and IC emission in Fig. \ref{spettro10000_ismthinfast}, where we give spectra at $\sim$ 10000 s after the burst. We evaluate the IC spectrum cutoff energy due to $\gamma \gamma$ interactions using Eq. \ref{cutoffism} for each set of model parameters, and display it with a thin vertical solid line. We find that in this case, pair attenuation has little to not effect on the predicted IC emission in the LAT energy band. The black dot-dashed line represents the LAT sensitivity obtained by applying the method explained in Sect. \ref{latsensitivity} for an integration time of $\sim$10000 s. We notice that with standard model parameters, LAT can detect afterglow IC emission for a sufficiently long integration time (see also Fig. \ref{thin_fast_Eb}). These spectra also show that as expected, the smaller the value of $\varepsilon_B$ is, which corresponds to greater values of the $(\epsilon_e / \epsilon_B)$ ratio, the higher the IC emission with respect to the synchrotron emission is (see Sect. \ref{contour}).  At early times IC emission dominates over synchrotron emission only at high energies, i.e., in the GeV band, but at late times IC emission dominates on synchrotron emission also at lower energies, i.e. in the hard X-ray/soft-$\gamma$-ray band \citep{corsi}. In Fig. \ref{thin_fast_Eb} we show how synchrotron (dashed line) and IC (solid line) light curves vary with $\epsilon_B$ and compare these light curves with the LAT threshold (black dot-dashed line).

If we change the model parameters so that during the first hundred seconds of ES emission the fireball is still in the slow cooling regime, in this case the 1 GeV IC flux is $ \sim 2 \times 10^{-12}~erg~cm^{-2}s^{-1}$ at around 10000 s after the burst, while the LAT sensitivity at this time is $\sim 9.4 \times 10^{-11}~erg~cm^{-2}s^{-1}$. Thus, in such a case, there is no possibility for the LAT to detect the IC component.


Now we discuss our predictions for a fireball expanding in a wind-like medium. In this case we require that the fireball remains in the fast cooling regime up to thousands of seconds after the burst, to maximize IC emission. We chose the model parameters according to the same criterion followed in the ISM case. During the fast cooling regime $Y$ depends only on the $(\epsilon_e/\epsilon_B)$ ratio and the IC flux increases with the fireball kinetic energy $E_{53}$ (see Eq. \ref{lumicfastwind}) as for the ISM case. Equation \ref{transitionwind} shows that the time the fireball remains in the fast cooling regime increases with the wind parameter $A_*$, and Eq. \ref{lumicfastwind} indicates that IC emission increases with $E_{53}$. In order to have a significant high energy emission we thus assume $E_{53}$=1 and $A_*=0.5$. In a wind the thin shell condition reads \citep{piro05}:

\begin{equation}
\label{thinwind}
\frac{T_2 A_* \Gamma_{0,2}^4}{E_{53}} < 0.06
\end{equation}

In this case we adopt $\Gamma_0=50$, so that with the parameters above the thin shell condition is verified for a prompt emission duration $\lesssim$ 200 s.

\begin{figure}[!htb]
\centering
\includegraphics[scale=0.35,angle=-90]{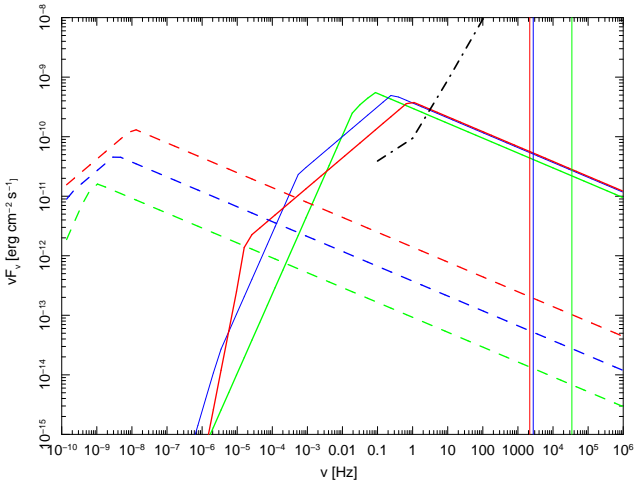}
\caption{Synchrotron (dashed lines) and IC (solid lines) spectra for a thin shell fireball expanding in a wind at $t \sim$ 10000 s. The model parameters are $\Gamma_0$=50, $E_{53}$=1, $A_*$=0.5, and $\epsilon_e$=0.2. Green curves correspond to $\epsilon_B=0.0001$, blue curves correspond to $\epsilon_B=0.001$, and red curves correspond to$\epsilon_B=0.01$. The thin solid lines indicate the cutoff energy due to pair production for the three IC spectra. The black dot-dashed line is the 5 $\sigma$ LAT sensitivity for an integration time of 10000 s.}
\label{spettro10000_windthinfast}
\end{figure}

\begin{figure}[!htb]
\centering
\includegraphics[scale=0.35,angle=-90]{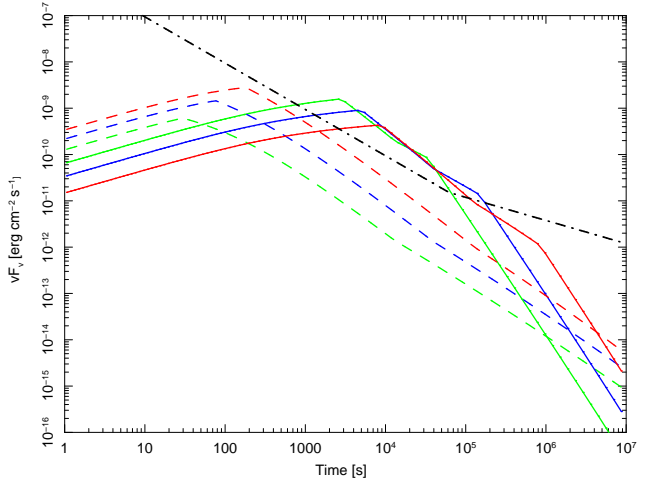}
\caption{Synchrotron (dashed line) and IC (solid line) light curves for a thin shell fireball expanding in a wind. The model parameters are the same as in Fig. \ref{spettro10000_windthinfast}. Synchrotron light curves are computed at $10^{18}$ Hz and IC light curves are computed at $2.4 \times 10^{23}$Hz. The green curves correspond to $\epsilon_B$=0.0001, the blue curves correspond to $\epsilon_B=$0.001 and the red curves correspond to $\epsilon_B$=0.01. The black dot-dashed line shows the 5$\sigma$ LAT threshold at 1 GeV as a function of the integration time.}
\label{thin_fastwind_Eb}
\end{figure}

In Fig. \ref{spettro10000_windthinfast} we show synchrotron and IC spectra as a function of $\epsilon_B$ at t $\sim$ 10000 s: as expected, the relative importance of IC and synchrotron emission increases with decreasing $\epsilon_B$ value, i.e with higher values of ($\epsilon_e/\epsilon_B$). Most of all this figure shows that, as for the ISM case, in this case LAT can detect afterglow IC emission if the integration time is sufficiently long (see also Fig. \ref{thin_fastwind_Eb}).

When the fireball is already in the slow cooling regime during the first phases of fireball evolution, as happens for a fireball expanding in an ISM LAT will again be unable to detect the IC component, even for sufficiently long integration times. In fact, the predicted 1 GeV IC flux at $\sim$ 10000 s is $\sim 3 \times 10^{-13}~erg cm^{-2}s^{-1}$, i.e. well below the corresponding LAT sensitivity.

In Sect. \ref{latsensitivity} we have shown that the GRID sensitivity is $\sim$ 15 times lower than that of the LAT when both the instruments are source dominated, and also lower for longer integration times (such as 10000 s) when the GRID is background dominated while the LAT not yet. As expected (see Sect. \ref{latsensitivity}) we find that at any time the predicted flux is below the AGILE sensitivity.


\section{GRB 940217 in the framework of the external shock model}
\label{940217}

\object{GRB 940217} was an energetic and long burst observed by the Energetic Gamma-Ray Experiment Telescope (EGRET) above 30 MeV with a duration of $\sim$ 5400 s and a possible emission of an 18 GeV photon around 4500 s \citep{hurley94}. The origin of this delayed high energy emission is still debated. At such high energies, IC is one of the best candidate emission mechanisms. We investigate whether IC emission from the afterglow of a ``standard '' thin shell fireball could explain the delayed high energy emission occurring in this event. \citet{hurley94} showed that above 30 MeV the best fit of the mean spectrum of the delayed high energy emission is a power law with photon index $\gamma=2.83 \pm 064$ (the 18 GeV photon is consistent with this fit at the 99\% confidence level). The integration of this power law gives a fluence $S=7\times 10^{-6}erg~ cm^{-2}$, which implies a mean flux $F \sim 1.4 \times 10^{-9}erg~ cm^{-2}s^{-1}$. Figure \ref{contourismlum} shows that a flux as high as that observed for \object{GRB 940217}, and greater also, is achieved for typical GRB model parameters in the case of thin shell fireballs when the fireball is in the fast cooling regime.

\begin{figure}[!htb]
\centering
\includegraphics[width=8.5cm,height=8.5cm,angle=90]{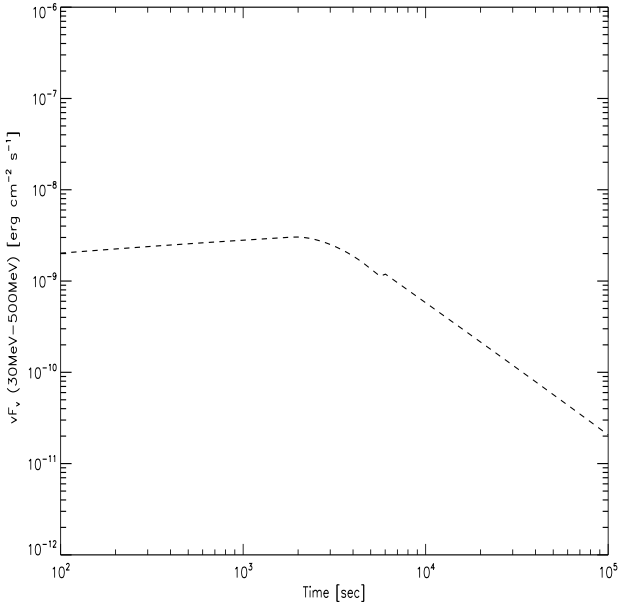}
\caption {IC light curve of \object{GRB 940217} in the 30 MeV-500 MeV energy range in the context of ES for a thin shell fireball expanding in an ISM. The model parameters are $E_{53}=5.0$, $n=3.0$, $\epsilon_e=0.07$,$\epsilon_B=0.001$, $p=2.5$, and the redshift is $z=1$. IC flux is roughly constant up to thousands of s, consistent with the observed delayed high energy emission.}
\label{lc940217}
\end{figure}

\begin{figure}[!htb]
\centering
\includegraphics[width=8.5cm,height=8.5cm,angle=-90]{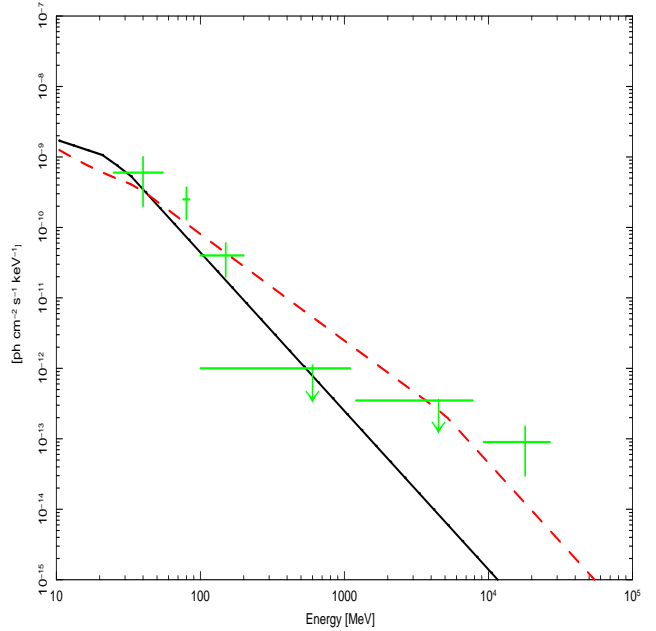}
\caption {IC plus synchrotron spectrum of \object{GRB 940217} at 500 s (red dashed line) and 5000 s (dark solid line) after the burst in the context of ES. The model parameters are the same as in Fig. \ref{lc940217}. Data points are taken from \citet{hurley94}.}
\label{spettro940217}
\end{figure}

During the fast cooling regime the photon index of the IC spectrum is $\gamma$= 3/2 for $\nu_{c,IC} < \nu_{obs} < \nu_{i,IC}$, and $\gamma=(\frac{p}{2}+1)$ for $\nu_{i,IC} < \nu_{obs}$. If, at the time of the delayed emission, $\nu_{i,IC}$ is below the observational band, for $p$=2.5 one obtains $\gamma$=2.25, which is consistent with the best fit value found by \citet{hurley94}. We thus require that at the time of the delayed emission $\nu_{i,IC} <$ 30 MeV. As noted by \citet{wei07}, the observational data indicate that the flux of the delayed emission is roughly constant with time. We succeed in finding a solution that accounts for both the spectral and temporal properties of the delayed high energy emission and its mean flux, as presented in  Fig.s \ref{lc940217} and \ref{spettro940217}. In particular in Fig. \ref{spettro940217} we compare the observed spectrum (green points) with that predicted in the framework of ES at $\sim$ 500 s (red-dashed line) and $\sim$ 5000 s (black solid line). We note that around 5000 s the predicted spectrum is fully consistent with the requirement that $\nu_{c,IC} < \nu_{i,IC} <$30 MeV. Around $500$ s the injection frequency $\nu_{i,IC}$ is still above 30 MeV, however the predicted spectrum is still in satisfactory agreement with the observed one. Below $\nu_{c,IC}$ the IC flux goes as $t^{1/3}$ and for $\nu_{c,IC} < \nu < \nu_{i,IC}$ it goes as $t^{1/8}$. This explains why the predicted flux is roughly constant along the delayed high energy emission timescale. However, we note that such a model is unable to account for the delayed 18 GeV photon.

\section{Prediction of GeV flares in the framework of DES}
\label{predizionihighen}
\subsection{X-ray flare properties}
\label{propflares}

We select from the sample of ''\emph{Gold}'' flares of \citet{falcone07} (i.e. the sample of X-ray flares with better statistics) a sub-sample of X-ray flares appearing in the first $\sim$ 1000 s after the burst and compare their spectral properties with those of the underling X-ray light curve. Flares and light curves are both fitted with a power law model. We retrieve the spectral properties of the X-ray continuum from \citet{willingale07}, who have performed the temporal and spectral analysis of a large sample of \emph{Swift} GRBs ($\sim 100$) in terms of two components, the first one attributed by them to the prompt $\gamma$-ray emission and the initial X-ray decay, and the second one thought to be due to the external shock, and developing in the afterglow. We list our sub-sample of flares in Table \ref{flare_spectra}, where we give the flare number, the end-time of the flare $t_{end}$ (as determined by \citet{falcone07}), the power law photon index of the flare $\Gamma_{flare}$, that of the prompt decay $\Gamma_{pd}$ and that of the afterglow decay $\Gamma_{ad}$, the time $T_a$ at which the power law afterglow decay starts, and the difference $\Delta \Gamma= \Gamma_{flare}- \Gamma_{ad}$ between the flare and the afterglow decay power law photon indices; all the quantities reported in this Table are at 90\% confidence level. We note that $T_a$ is always larger than the time of the flare appearance. We find that for 16 flares out of 28, the flare photon index is not consistent with that of the prompt decay, nor with that of the afterglow decay. For 5 out of 28 flares $\Gamma_{flare}$ is consistent with both $\Gamma_{pd}$ and $\Gamma_{ad}$. For one flare $\Gamma_{flare}$  is consistent only with $\Gamma_{pd}$, and for 5 out of 28 flares $\Gamma_{flare}$ is consistent only with $\Gamma_{ad}$. In total, there are 10 out of 28 flares for which the spectrum is consistent with that of the second component of GRBs light curve (that can be ascribed to the external shock). We also note that in the majority of cases the spectral difference between the flare and the afterglow decay spectrum is $\Delta \Gamma \sim$0.5. Such hard-to-soft spectral evolution can be easily accounted for in the context of the ES model. Indeed, during the fast cooling regime, when the injection frequency $\nu_m$ passes into the observational band, there is a hard-to-soft spectral evolution of $\Delta \Gamma=(p-1)/2$, with $0.5 \lesssim \Delta \Gamma \lesssim 1$ for $2.0 \lesssim p \lesssim 3.0$. Similarly, during the slow cooling regime, when the cooling frequency $\nu_c$ passes into the observational band there is an hard-to-soft spectral evolution of $\Delta \Gamma=$0.5. If the fireball expands in a wind-like medium, the cooling frequency increases with time, thus one can also explain a soft-to-hard spectral evolution of $\Delta \Gamma=$0.5, that was observed in some cases (see Table \ref{flare_spectra}).

\begin{table*}[!htb]
\caption{Properties of power law spectral fits to the Gold sample of flares of \citet{falcone07} and to the total X-ray light curve. The spectral properties of the global X-ray light curve are taken from \citet{willingale07}. Errors are at 90\% confidence level. }
\centering
\label{flare_spectra}
\begin{tabular}{c c c c c c c c} 
\hline 
 GRB    & flare & $t_{end}$ & $\Gamma_{flare}$ & $\Gamma_{pd}$ & $T_a$  & $\Gamma_{ad}$ & $\Delta \Gamma= \Gamma_{flare}- \Gamma_{ad}$\\ 
        &       &   [s]     &                        &                  &        [s]                  &                &    \\
\hline \hline   
050219A &  1  & 453 & $2.67^{+0.41}_{-0.34}$ & $2.02 \pm 0.20$ & $30900^{646000}_{3090}$  &  $1.89 \pm 0.24$  & $0.78^{+0.84}_{-0.42}$ \\
\hline
050607  &  2  & 640 & $2.40^{+0.23}_{-0.20}$ & $1.77 \pm 0.48$  & $20900^{64600}_{6310}$  & $1.74 \pm 0.18$   & $0.66^{+0.29}_{-0.27}$ \\
\hline
050712  &  1  & 564 & $2.13^{+0.18}_{-0.17}$ & $1.91 \pm 0.06$  & $67600^{173800}_{20000}$ & $1.80 \pm 0.26$  & $0.33^{+0.32}_{-0.31}$ \\
\hline
050713  &  1  & 155 & $2.19 \pm 0.11$        & $2.30 \pm 0.07$  & $12300^{19500}_{7940}$   & $1.86 \pm 0.22$  & $0.33 \pm 0.25$ \\

        &  2  & 210 & $3.30^{+0.68}_{-0.51}$ & $2.30 \pm 0.07$  & $12300^{19500}_{7940}$   & $1.86 \pm 0.22$  & $1.44^{+0.72}_{-0.56}$ \\
\hline
050716  &  1 & 211 & $1.22^{+0.93}_{-0.75}$ & $1.33 \pm 0.03$  & $17400^{758600}_{1740}$  & $1.93 \pm 0.14$  & $-0.71^{+0.94}_{-0.76}$ \\
        &  2  & 483 & $3.38^{+0.93}_{-0.63}$ & $1.33 \pm 0.03$  & $17400^{758600}_{1740}$  & $1.93 \pm 0.14$  & $1.45^{+0.94}_{-0.65}$ \\
\hline
050724  &  1    &   230     & $1.77 \pm 0.03$        & $1.95 \pm 0.07$  & $105000^{10^6}_{31620}$  & -        & - \\
        &  2    &   342     & $2.94^{+0.13}_{-0.12}$ & $1.95 \pm 0.07$  & $105000^{10^6}_{31620}$  & -        & - \\
\hline
050726  &  2    &   324     & $2.55^{+0.68}_{-0.50}$ & $1.94 \pm 0.07$  &  -               & $1.93 \pm 0.27$ & $0.62^{+0.73}_{-0.57}$ \\
\hline
050730  &  1  & 280 & $1.71 \pm 0.12$        & $1.33 \pm 0.08$  & $13500^{16200}_{11800}$  & $1.62 \pm 0.05$ & $0.09 \pm 0.13$ \\
        &  2  & 611 & $1.66 \pm 0.05$        & $1.33 \pm 0.08$  & $13500^{16200}_{11800}$  & $1.62 \pm 0.05$ & $0.04 \pm 0.07$  \\
        &  3  & 795 & $2.20^{+0.14}_{-0.13}$ & $1.33 \pm 0.08$  & $13500^{16200}_{11800}$  & $1.62 \pm 0.05$ & $0.58^{+0.15}_{-0.14}$ \\
\hline
050802  & 1   & 457 & $2.13^{+0.46}_{-0.36}$ & $1.91 \pm 0.19$  & $9120^{11200}_{7240}$    & $1.81 \pm 0.09$ & $0.32^{+0.47}_{-0.37}$ \\
\hline
050820A & 1   & 382 & $0.82 \pm 0.04$        & $1.87 \pm 0.09$  & $9120^{10960}_{6920}$    & $1.74 \pm 0.75$ & $-0.92 \pm 0.75$ \\
\hline
050822  &  1  & 190 & $1.78^{+0.30}_{-0.27}$ & $2.60 \pm 0.06$  & $25700^{37150}_{17000}$  & $2.13 \pm 0.10$ & $-0.35^{+0.32}_{-0.29}$ \\
        &  2  & 276 & $2.86^{+0.27}_{-0.24}$ & $2.60 \pm 0.06$  & $25700^{37150}_{17000}$  & $2.13 \pm 0.10$ & $0.73^{+0.29}_{-0.26}$ \\
        &  3  & 758 & $4.36^{+1.45}_{-1.03}$ & $2.60 \pm 0.06$  & $25700^{37150}_{17000}$  & $2.13 \pm 0.10$ & $2.23^{+1.45}_{-1.04}$  \\
\hline
050904  &  1  & 570 & $1.78 \pm 0.09$        & $1.44 \pm 0.04$  & $10500^{33000}_{1510}$   & $2.00 \pm 0.14$ & $-0.22 \pm 0.17$ \\
\hline
050922B &  1  & 280 & $3.94^{+0.78}_{-0.58}$ & $2.64 \pm 0.08$  & $209000^{389000}_{93300}$ & $2.33 \pm 0.25$ & $1.61^{+0.82}_{-0.63}$ \\
        &  2  & 611 & $2.66^{+0.86}_{-0.60}$ & $2.64 \pm 0.08$  & $209000^{389000}_{93300}$ & $2.33 \pm 0.25$ & $0.33^{+0.90}_{-0.65}$ \\
        &  3  & 795 & $2.36 \pm 0.10$        & $2.64 \pm 0.08$  & $209000^{389000}_{93300}$ & $2.33 \pm 0.25$ & $0.03 \pm 0.27$ \\
\hline
051227  &  1  & 245 & $1.53^{+0.15}_{-0.14}$ & $1.41 \pm 0.22$  &  -                      & $1.58 \pm 0.40$  & $-0.05^{+0.43}_{-0.42}$ \\
\hline
060111A &  1  & 196 & $2.89^{+0.14}_{-0.13}$ & $2.35 \pm 0.04$  & $2140^{562000}_{214}$  & $2.39 \pm 0.12$ & $0.50 \pm 0.18$ \\
        &  2  & 203 & $2.86^{+0.18}_{-0.17}$ & $2.35 \pm 0.04$  & $2140^{562000}_{214}$  & $2.39 \pm 0.12$ & $0.47^{+0.22}_{-0.21}$ \\
        &  3  & 433 & $2.27 \pm 0.05$        & $2.35 \pm 0.04$  & $2140^{562000}_{214}$  & $2.39 \pm 0.12$ & $-0.12 \pm 0.13$ \\
\hline
060124  &  1  & 644 & $1.21 \pm 0.01$ & $1.86 \pm 0.11$   & $39800^{49000}_{29500}$  & $2.28 \pm 0.11$ & $-1.07 \pm 0.11$ \\
        &  2  & 1007 & $1.67 \pm 0.02$ & $1.86 \pm 0.11$   &  $2140^{49000}_{29500}$  & $2.28 \pm 0.11$ & $-0.12 \pm 0.11$ \\
\hline
\end{tabular}
\end{table*}


Another important quantity is the fireball kinetic energy $E_{53}$ that significantly affects the predicted synchrotron and IC flare emission (see e.g. Eqs. \ref{lumicfastvalue} and \ref{lumicfastwind}); this quantity can be estimated from the emitted isotropic energy $E_{iso}$ of the burst correcting it for the fireball radiative efficiency $\eta$. In Sect. \ref{thin} we have shown that a favorable condition for a detectable IC high energy emission is to have a burst with $E_{53} \gtrsim$1. It is thus important to understand what fraction of GRBs satisfy this energy requirement. To this end we have further selected from \citet{falcone07} a sub-sample of \emph{Swfit} GRBs with known redshift, and report them in Table \ref{flare_properties}. When it is possible we take the value of $E_{iso}$ directly from \citet{amati07}. For the other bursts we estimate $E_{iso}$ according to the following method. The burst isotropic energy $E_{iso}$ is given by the integral of the burst spectral model in the 1-$10^4$ keV band (see e.g \citet{amati02}). In this energy band the spectrum of a burst is typically described by a Band model \citep{band93}, thus in order to calculate this integral one needs to know the two power law photon indeces and the peak energy of the Band function. The only information we can retrieve from the on-line \emph{Swift} GRBs table \footnote{http://swift.gsfc.nasa.gov/docs/swift/archive/grb\_table.html/} are the 15-150 keV fluence $S$ and the prompt emission photon index $\Gamma$ of the burst measured by the Burst Alert Telescope (BAT). We therefore run a set of integrations by varying $E_{peak}$ (for given values of spectral indices, see below) and derive the corresponding $E_{iso}$ using the 15-150 keV BAT fluence to normalize the spectrum. The chosen values of $E_{peak}$ and $E_{iso}$ are those satisfying the Amati relationship \citep{amati02,amati06}. In each integration, depending on the value of $E_{peak}$, we identify the observed power law photon index with one of the two indices of the Band function, fixing the other one to a canonical value (-1 and -2 for the low and the high energy power laws respectively). We find that $\sim$ 30 \% of the bursts of our sample have isotropic energy $E_{iso} \gtrsim 10^{53}~erg$ (see Table \ref{flare_properties}). Finally we assume a radiative efficiency $\eta \sim 0.1$, which implies that $\sim$ 30 \% of the GRBs with flares in their X-ray light curve have fireball kinetic energy $\gtrsim 10^{54}$ erg.

\begin{table*}[!htb]
\caption{ Sample of \emph{Swfit} GRBs with known redshift having flares in their light curve. We report in this table their prompt emission properties, and in particular their isotropic energy $E_{iso}$. When it is possible we take the value of $E_{iso}$ directly from \citet{amati06}; we mark these bursts with $^*$. For the other bursts we assume the validity of the Amati relation, and determine the values of the intrinsic burst peak energy $E_{peak}$ and $E_{iso}$ implied by the observed fluence of the burst itself. }
\centering
\label{flare_properties}
\begin{tabular}{c c c c c c c c c c} 
\hline 
 GRB        & redshift & S (15-150 keV)     & $D_l$          & $\Gamma$ &  $E_{peak}$   &  $E_{iso}$     \\ 
            &  z & [$ 10^{-7} erg~cm^{-2}$] & [$10^{28}~cm$] &          &   [keV]       & [$10^{53}~erg$]\\
\hline \hline   
050406      & 2.44  &   0.806             &   5.94         &    2.44  &   23.0  &   0.012  \\
050724$^*$  & 0.26  &   11.8              & 0.395          &  2.19    &  -         & $3.0 \times 10^{-3}$ \\
050730$^*$  & 3.97  &   24.2              & 10.66          &   1.52   & -             & 2.6    \\
050802      & 1.71  &  22.0               & 3.846          & 1.55     &  158.0  & 0.35    \\
050803      & 0.42  &       22.3          & 0.69           & 1.40     &  72.0   & 0.09    \\
050814      & 5.3   &        18.3         &  15.0          & 1.86     &  165.0  & 0.38    \\
050820A     & 2.61  &       40.1          & 6.447          & 1.21     &  300.0  & 1.1   \\
050904$^*$  & 6.3   &          50.7       & 18.3           & 1.30     &   -           & 19.3   \\
050908      & 3.35  &          4.91       & 8.7            & 1.86     &  68.0   & 0.08    \\
060108      & 2.03  &         3.7         & 4.75           & 2.01     &   42.0   & 0.035   \\
060115$^*$  & 3.53  &         18.0        & 9.27           &  1.0     &     -         & 0.91    \\
06012$^*$4  & 2.3   &        4.6          &  5.6           &  1.89    &     -         & 4.8     \\
\hline     
\end{tabular}
\end{table*}


\subsection{Inverse Compton and synchrotron emission in the delayed external shock scenario: X-ray and high energy flares}
\label{thick}

In the preceding section we have summarized some of the more relevant flare properties and have shown that there are flares for which temporal-spectral properties can be potentially explained in the framework of external shocks, and more specifically, in the framework of the DES scenario (i.e. thick shell fireballs, \citep{sari99}). In particular, we have shown that a DES can also potentially explain spectral variations of the order of 0.5-1.0, depending on the radiative regime and electron population spectral index. In the DES scenario the duration of the central engine activity $t_{eng}$ is larger than the time $t_{dec}$ when the fireball starts to decelerate, and most of the energy is transferred to the surrounding material at late times, around the end of the engine activity. The flare would thus be produced by an ES caused by a  central engine which remains active until the time of the flare occurrence (see \citet{piro05,galli06} for details). We also stress that in this context, with the flare being ascribed to the onset of afterglow emission, only one flare can be explained. To explain multiple X-ray flares other mechanisms, such us LIS by a long duration central engine, refreshed shock, or ES with a clumpy medium, are required. An important implication of models proposed to explain the origin of flares is that X-ray flare photons can be IC scattered, thus producing high energy (MeV to TeV) flares. In particular, in the context of DES the X-ray flare photons are IC scattered by the afterglow electrons, and X-ray and high energy flares are produced by the same source region and electron population.

Before studying in detail the properties of the predicted IC emission associated with flares, we first provide a rough estimation of the mean flux and the peak energy of the high energy flare predicted by the DES scenario as a function of X-ray flare flux and peak energy. This allows a first-order prediction of the detectability of high energy flares based solely on their lower energy properties. The IC flux can be estimated from that observed in X-ray through the following general equation (see \citet{esin01}):
 
\begin{equation}
\label{stimaIC}
\nu_{p,IC}F_{\nu,IC}=Y\nu_{p}F_{\nu}
\end{equation}

where $\nu_{p,IC}$ and $\nu_{p}$ are the peak frequencies of the IC and synchrotron components with $F_{\nu,IC}$ and $F_{\nu}$ the related peak flux densities, and $Y$ is given by Eq. \ref{eq:lumic}. The equation above tells us that the flux, and consequently also the fluence, associated with the IC component is roughly $Y$ times those of the X-ray component. \citet{falcone07} have derived the mean (0.2-10.0) keV flare fluence to be $S \approx 3 \times 10^{-7}~erg~ cm^{-2}$ and the typical flare duration to be few hundred seconds for flares appearing within $\sim$ 1000 s after the burst. This implies a mean X-ray flare flux of $\approx 3 \times 10^{-9}~erg~cm^{-2}s^{-1}$, and a mean 1 keV X-ray flux density of $\approx 0.5$ mJy for a typical flare spectrum. On the other hand, the Compton factor $Y$ strongly depends on the values of the parameters $\epsilon_e$ and $\epsilon_B$. For example, at equipartition $Y$ is of the order of unity, and in this case we thus expect that the fluence of the high energy flare is of the same order of magnitude as that of the X-ray flare, i.e. $\approx 3 \times 10^{-7}~erg cm^{-2}$. Such a fluence is below the 5$\sigma$ 1 GeV LAT fluence threshold, $\sim 10^{-6}~erg~cm^{-2}$, and is of the same order of magnitude as the 5$\sigma$ 100 MeV LAT fluence threshold, $\sim 2.7 \sim 10^{-7}~erg~cm^{-2}$ (with the LAT on-axis 100 MeV effective area $\sim$ 3000 $cm^2$; see Sect. \ref{latsensitivity} for the estimation of the LAT sensitivity). When $Y$ is above unity (this is common during the fast cooling regime for typical model parameters; see Fig. \ref{contourism}), the fluence carried by the high energy flare can be larger than that carried by the X-ray flare, and the detection of high energy flares improves. It is important to note that the detection of high energy flares strongly depends on the peak energy with respect to the LAT energy band.

In Sect. \ref{comparison} we have shown that we expect similar fluxes for a fireball expanding in an ISM and in a wind-like medium. Therefore, in this section we focus our attention only on the case of a thick shell fireball expanding in an ISM. We now estimate the peak energy of the IC component as a function of the observed X-ray flare flux and peak energy. To this end we first determine the values of the model parameters that satisfy the observed flare properties in the context of the DES scenario. We reasonably assume that at the time of the flare appearance the fireball is in the fast cooling regime, and that the injection frequency $\nu_m$ (i.e. the flare peak frequency in the $\nu-\nu F_{\nu}$ space) is below the observational frequency $\nu_{obs}$. Under these assumptions, if we fix the fireball kinetic energy to $E_{53}=10$ (we have shown in the preceding section that this is a reasonable choice), $p=2.5$ and $z=1.0$, the flare mean X-ray flux and peak frequency $\nu_m$ depend only on the two parameters $\epsilon_e$ and $\epsilon_B$, and we can search for a pair of $\epsilon_e$ and $\epsilon_B$ values accounting for the observed flare properties. When $\nu_{obs} > \nu_{m}$, the synchrotron flux density is given by $F_{\nu}=F_{\nu_c}(\nu_{obs}/\nu_m)^{-p/2}(\nu_c/\nu_m)^{1/2}$. In this equation we substitute $F_{\nu_c}$, $\nu_m$ and $\nu_c$ using Eqs. (64), (22) and (27) of \citet{panaitescu00}, respectively. We thus find two equations relating $\epsilon_e$ and $\epsilon_B$ to the 1 keV X-ray flare flux density $F_{1keV}$ and peak frequency $\nu_m$, and the time $t_{obs}$ when we measure the flare properties:
 
\begin{equation}
\label{par1}
\epsilon_e \simeq 0.017 \biggl( \frac{E_{iso,53}}{\eta_{,-1}} \biggr)^{1/3} \biggl( \frac{\nu_m}{1 keV} \biggr)^{5/6} \biggl( \frac{F_{1keV}}{1 mJy} \biggr)^{-2/3} t_{obs}^{1/3}
\end{equation}

\begin{equation}
\label{par2}
\epsilon_B \simeq 1.3 \times 10^{-6} \biggl( \frac{E_{iso,53}}{\eta_{,-1}} \biggr)^{-7/3} \biggl( \frac{\nu_m}{1 keV} \biggr)^{-4/3} \biggl( \frac{F_{1keV}}{1 mJy} \biggr)^{8/3} t_{obs}^{5/3}
\end{equation}

where $E_{iso,53}$ is the emitted isotropic energy in unity of $10^{53}$ erg, and $\eta_{,-1}$ is the fireball radiative efficiency in unity of 0.1. The ratio $E_{iso,53} / \eta$ is the way we estimate the fireball kinetic energy $E_{53}$. At this point we can finally write the Compton factor $Y$, the peak energy of the high energy flare $\nu_{p,IC}$ and its flux $\nu_{p,IC}F_{\nu,IC}$, in terms of only $F_{1~keV}$, and the flare peak energy $\nu_m$ and peak flux density $F_{\nu_m}$:

\begin{equation}
\label{comptonfactor}
Y \sim 53 \biggl( \frac{E_{iso,53}}{\eta_{,-1}}\biggr)^{4/3} \biggl( \frac{\nu_m}{1 keV} \biggr)^{13/12} \biggl( \frac{F_{1keV}}{1 mJy} \biggr)^{-5/3} t_{obs}^{-2/3}
\end{equation}

\begin{equation}
\label{piccohighflare}
\nu_{p,IC} \simeq 396 n_{0}^{-1/4} \biggl( \frac{E_{iso,53}}{\eta_{,-1}}\biggr)^{11/12} \biggl( \frac{\nu_m}{1 keV} \biggr)^{8/3} \biggl( \frac{F_{1keV}}{1mJy} \biggr)^{-4/3} t_{obs}^{-1/12} ~GeV
\end{equation}

\begin{equation}
\begin{split}
\label{flussohighflare}
\nu_{p,IC} F_{\nu,IC} & = Y F_{\nu_m} \nu_m \sim 1.26 \times 10^{-7} \biggl( \frac{E_{iso,53}}{\eta_{,-1}}\biggr)^{4/3} \biggl( \frac{\nu_m}{1 keV} \biggr)^{25/12} \\
          & \quad \biggr( \frac{F_{1keV}}{1 mJy} \biggl)^{-5/3} \biggr( \frac{F_{\nu_m}}{1 mJy} \biggl) t_{obs}^{-2/3}~erg~cm^{-2}s^{-1}
\end{split}
\end{equation}

The analytical equations given above hold only during the fireball deceleration phase. We thus take $t_{obs}$=200 s in our calculations, i.e. $t_{obs}$ is larger than the fireball deceleration time (which corresponds to the flare maximum in the DES scenario), but sufficiently close the peak time of the flare. From the preceding Eqs. we find that if we observe an X-ray flare with 1 keV flux density $F_{1keV}$=0.5 mJy and peak frequency $\nu_m$=10 eV, and the density of the uniform interstellar medium is n=10, for $t_{obs} \sim$ 200 s then $Y \sim$1. We also expect a high energy flare with peak flux $\sim 3 \times 10^{-10}~erg~cm^{-2}~s^{-1}$ and peak frequency $\nu_{p,IC} \sim$ 25 MeV. If the X-ray flare peak frequency is $\nu_m$=1 keV and the other quantities do not vary, then $Y \sim$100 and we expect a high energy flare with peak flux $\sim 10^{-7}~erg~cm^{-2}~s^{-1}$ and peak frequency $\nu_{p,IC} \sim$ 5 TeV. 

We now present two numerical solutions corresponding to two flares with 1 keV flux density of $\sim$ 0.5 mJy and peak energies around 10 eV and 1 keV, respectively. We take the density of the external medium $n=10$, the initial fireball Lorentz factor $\Gamma_0=150$, and values of $\epsilon_e$ and $\epsilon_B$ that satisfy the conditions of fast cooling regime, Thompson regime and thick shell fireball (which can be obtained reversing Eq. \ref{thinism}). We assume a typical value of 500 s for the time of the flare occurrence \citep{chincarini07,falcone07} and apply the thick shell condition, shifting the origin of time to this instant \citep{piro05,galli06}, $t_0=500$ s.

\begin{figure}[!htb]
\centering
\includegraphics[scale=0.37,angle=-90]{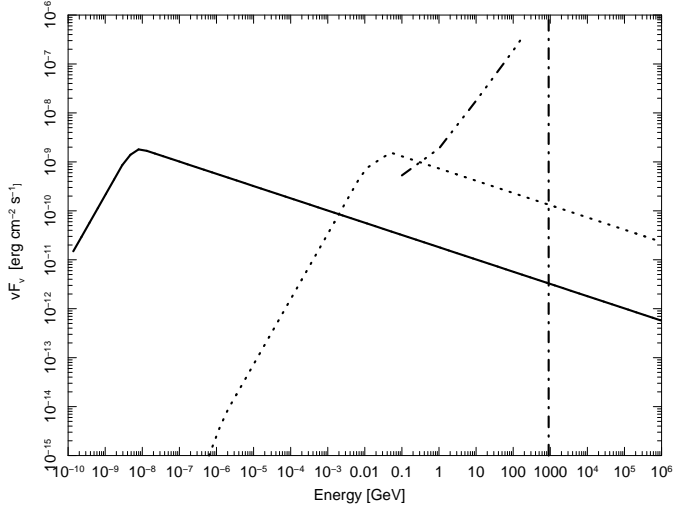}
\caption{Synchrotron (solid line) and IC (dotted line) spectra for a thick shell fireball expanding in an ISM $\sim$ 200 s after the start of the flare, $t_0 \sim 500$ s. The model parameters are $E_{53}$=10.0, $\Gamma_0$=150, $n$=10.0, $\epsilon_e$=0.01, $\epsilon_B=10^{-3}$, $p=2.5$ and $z=1.0$. With this set of model parameters the fireball is in the fast cooling regime at the time of the flare occurrence, and the mean flare peak energy is around 10 eV. The dot-dot-dot-dashed line is the 5$\sigma$ LAT sensitivity for an integration time of 500 s. The vertical dot-dashed line indicates the cutoff energy due to pair production for the IC spectrum. }
\label{spettro200s_ismthickfast_vp10ev}
\end{figure}

\begin{figure}[!htb]
\centering
\includegraphics[scale=0.37,angle=-90]{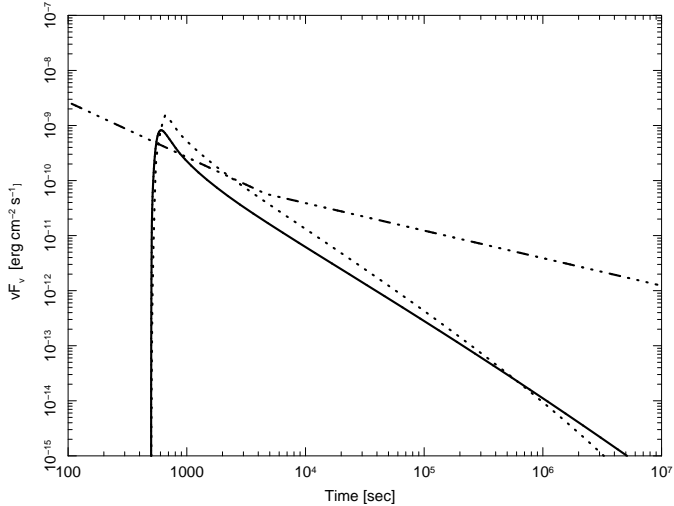}
\caption{Synchrotron (solid line) and IC light curves at 100 MeV (dotted line) for a fireball expanding in an ISM in the framework of DES. The model parameters are the same as in Fig. \ref{spettro200s_ismthickfast_vp10ev}. The origin of the time is shifted to $t_0$=500 s. With this set of model parameters the fireball is in the fast cooling regime at the time of the flare occurrence, and the mean flare peak energy is around 10 eV. The dot-dot-dot-dashed curves shows the 5$\sigma$ 100 MeV LAT sensitivity as a function of the integration time.}
\label{thickism_lc_vp10ev}
\end{figure}

We present in Figs. \ref{spettro200s_ismthickfast_vp10ev} and \ref{thickism_lc_vp10ev} a numerical solution corresponding to a flare peak energy of $\sim$ 10 eV. Figure \ref{spettro200s_ismthickfast_vp10ev} shows synchrotron (solid line) and IC (dotted line) spectra around 200 s after the start of the flare; the predicted high energy flare can be detected by the LAT up to $\sim~1$ GeV. In Fig. \ref{thickism_lc_vp10ev} we give synchrotron (1 keV; solid line) and IC (100 MeV; dotted line) light curves, and compare the latter with the 100 MeV LAT sensitivity (dot-dot-dot-dashed line). When the peak energy of the X-ray flare is around 10 eV the peak energy of the high energy flare is around 200 MeV and there is a good temporal correlation between the X-ray and high energy flares; this is expected in the context of the DES scenario, when the peak energies of X-ray and high energy flares are below or very close to the relative observational bands.

\begin{figure}[!htb]
\centering
\includegraphics[scale=0.37,angle=-90]{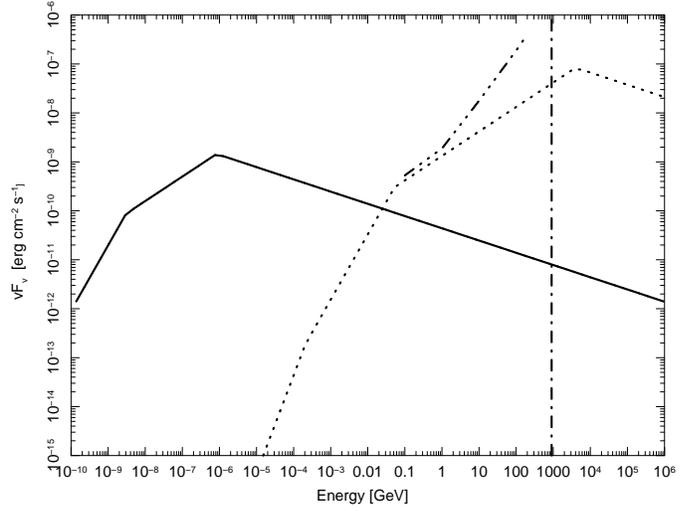}
\caption{Synchrotron (solid line) and IC (dotted line) spectra for a thick shell fireball expanding in an ISM $\sim$ 200 s after the start of the flare, $t_0=500$ s. The model parameters are $E_{53}$=10.0, $\Gamma_0$=150, $n$=10.0, $\epsilon_e$=0.27, $\epsilon_B=10^{-5}$, $p=2.5$  and $z=1.0$. With this set of model parameters the fireball is in the fast cooling regime at the time of the flare occurrence, and the mean flare peak energy is around 1 keV. The dot-dot-dot-dashed line is the 5$\sigma$ LAT sensitivity for an integration time of 500 s. The vertical dot-dashed line indicates the cutoff energy due to pair production for the IC spectrum. }
\label{spettro200s_ismthickfast_vp1kev}
\end{figure}

\begin{figure}[!htb]
\centering
\includegraphics[scale=0.37,angle=-90]{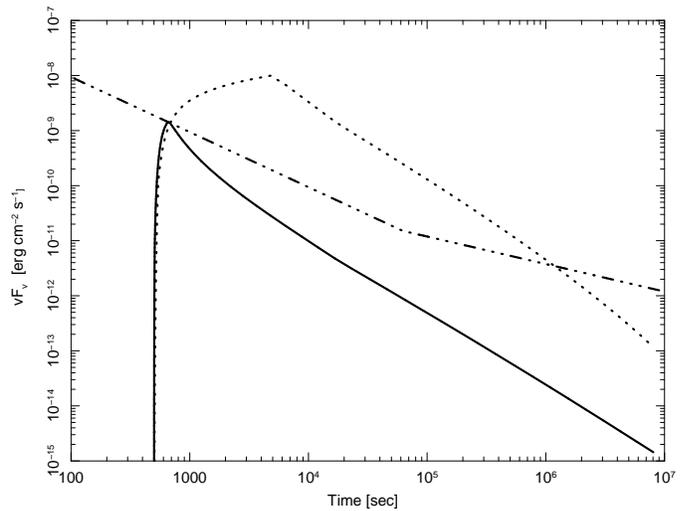}
\caption{Synchrotron (solid line) and IC light curves at 1 GeV (dotted line) for a fireball expanding in an ISM in the framework of DES. The model parameters are the same as in Fig. \ref{spettro200s_ismthickfast_vp1kev}. The origin of the time is shifted to $t_0$=500 s. With this set of model parameters the fireball is in the fast cooling regime at the time of the flare occurrence, and the mean flare peak energy is around 1 keV. The dot-dot-dot-dashed curve shows the 5$\sigma$ 1 GeV LAT sensitivity as a function of the integration time.}
\label{thickism_lc_vp1kev}
\end{figure}

Fig.s \ref{spettro200s_ismthickfast_vp1kev} and \ref{thickism_lc_vp1kev} present a numerical solution corresponding to an X-ray flare peak energy of $\sim$ 1 keV. In this case the $\nu F_{\nu}$ peak flux of the IC component is roughly two orders of magnitude larger than the synchrotron one; this is a good condition for the detection of high energy flares. In Fig. \ref{spettro200s_ismthickfast_vp1kev} we show synchrotron (solid line) and IC (dotted line) spectra $\sim$ 200 s after the start of the flare, and in Fig. \ref{thickism_lc_vp10ev} we give synchrotron (1 keV; solid line) and IC (1 GeV; dotted line) light curves, and compare the latter with the 1 GeV LAT sensitivity (dot-dot-dot-dashed line). The fluence carried by the high energy flare is significantly larger than that carried by the X-ray flare, and integrating for a sufficiently long time, LAT can easily detect the expected high energy flare. However, in such a case the peak energy of the high energy flare is of the order of TeV (see Fig. \ref{spettro200s_ismthickfast_vp1kev})), thus there is a temporal delay between the X-ray and the high energy flare in the LAT energy band.

\section{Individual cases: GRB 011121 and XRF 011030}
\label{varicasi}

In this section we estimate the IC component in the framework of DES for \object{XRF 011030} and \object{GRB 011121}. In the case of \object{XRF 011030}, \citet{galli06} have shown that a late X-ray flare occurs in this event around 1300 s and it can be explained in the context of the DES scenario, both for a fireball expanding in a wind environment and in an ISM. However, in their computation, \citet{galli06} did not account for the cooling due to the presence of an IC component, thus now we want to assess whether the presence of this component affects their findings.

In the ISM case, \citet{galli06} found a broadband (radio-to-X-ray) solution for a jetted fireball with $E_{53}=0.03$, $\Gamma_0=130$, $n=5$, $\varepsilon_e=0.29$, $\varepsilon_B=8\cdot10^{-5}$, $p=2.1$, $z=1$ and break time $T_b=8 \cdot 10^5$ sec. This solution implies that during the fast cooling regime the relative importance of IC and synchrotron components is $Y \sim 27$, thus the X-ray synchrotron flux is suppressed by a factor $(1+Y) \sim 30$. On the contrary, with this set of model parameters, the optical and radio bands are below the cooling frequency $\nu_c$ at the time of observations, thus in these bands the synchrotron flux is not affected by the presence of IC emission. To increase the predicted X-ray flux in order to be consistent with the observations, one should increase the fireball energy $E_{53}$ and/or $\epsilon_B$, but this has the effect of also increasing the predicted optical and radio fluxes bringing them above the observed fluxes (see Eqs. in appendix B of \citet{panaitescu00}). We thus conclude that in presence of IC emission for a fireball expanding in an ISM, we cannot find a broadband solution.

\begin{figure}[!htb]
\centering
\includegraphics[scale=0.35,angle=-90]{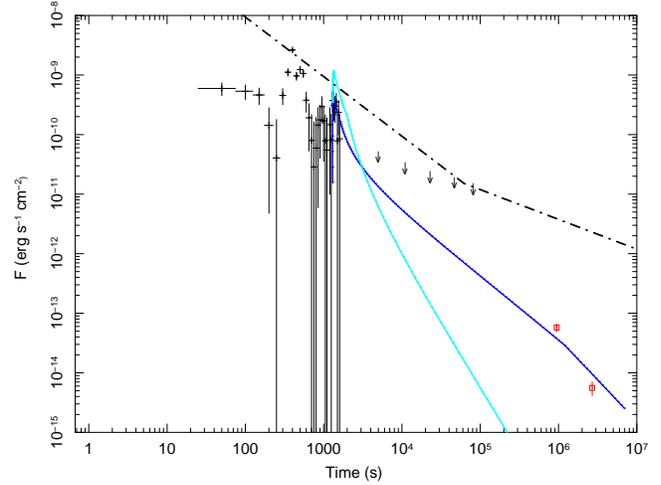}
\caption{Synchrotron (blue) and 1 GeV IC (light blue) light curves for XRF 011030 in the case of a thick shell fireball expanding in a wind density profile. The model parameters are $E_{53}$=0.3, $\Gamma_0$=70, $A_*$=0.055, $\epsilon_e$=0.02, $\epsilon_B$=0.001, and $p$=2.1. The black dot-dashed line is the LAT sensitivity at 1 GeV as a function of the integration time.}
\label{011030_wind_ic}
\end{figure}

In the wind case, \citet{galli06} found a broadband solution with $E_{53}$ = 0.3, $\Gamma_0$=60, $A_*$=0.055, $\epsilon_e$=0.02, $\epsilon_B$=0.001, z=1 and p=2.1. With such parameters, in presence of an IC component the X-ray synchrotron flux is reduced at maximum by a factor $\sim 3$ during the fast cooling regime. At the time of the flare the fireball passes from the fast to the slow cooling regime, thus $\nu_x > \nu_c > \nu_i$ and $F_{\nu,x} \propto \nu_x^{-p/2} \nu_i^{(p-1)/2} \nu_c^{1/2} \propto (1+Y)^{-1}$. In this case the cooling frequency increases with time, thus around $10^6$ s $\nu_c > \nu_x$ and $F_{\nu,x}$ does not depend on $\nu_c$. Finally, optical and radio emission are also not affected by the presence of an IC component. This means that in this case, only the X-ray light curve around the time of the flare occurrence is affected by the presence of IC emission. As shown in \citet{galli06}, in this phase the emission strongly depends on the initial fireball Lorentz factor $\Gamma_0$, thus we can find a new broadband solution simply by increasing this parameter from $\Gamma_0=60$ to $\Gamma_0=70$ (see Fig. \ref{011030_wind_ic}). The presence of an IC component also has the effect of reducing the cooling frequency $\nu_c$ by a factor $(1+Y)^2$, thus causing a delay of the time when it crosses the X-ray band. However, the spectral break related to the crossing of $\nu_c$ in the X-ray band is still consistent with that suggested by the two late time CHANDRA observations (red squares in Fig. \ref{011030_wind_ic}). In Fig. \ref{011030_wind_ic} we show the predicted synchrotron (blue line) and 1 GeV IC (light blue line) light curves. As expected from the ES model, there is a good temporal correlation between the X-ray and GeV flare. Indeed, at the time of the X-ray flare the peak of IC component is around 1 GeV, and this also ensures that the presence of an IC component does not affect the spectrum of the X-ray flare. We also note that even if \object{XRF 011030} is not a bright burst, the predicted IC emission related to its X-ray flare is potentially detectable by the LAT.

Another interesting case is that of \object{GRB 011121}. The broadband afterglow data of \object{GRB 011121} and its X-ray flare were explained by \citet{piro05} and \citet{galli06} by a thick shell fireball expanding in a wind-like medium, with efficiencies $\epsilon_e$=0.01 and $\epsilon_B=$0.5. As we have showed in Sect. \ref{contour}, IC emission can dominate over synchrotron emission only if $\epsilon_e > \epsilon_B$. This means that even if the X-ray flare occurring in \object{GRB 011121} is very bright, a non significant contribution of IC emission is expected for this event. 

\section{Discussion and conclusions}
\label{conclusionigen}

In this paper we study the relative importance of synchrotron and IC emission in the framework of ES for a fireball expanding in an ISM and in a medium with a wind density profile. We first focus our attention on the ''standard'' case of a thin shell fireball, and study synchrotron and IC emission as a function of the model parameters that mainly determine the relative importance of these two emission mechanisms. We find that the importance of the IC component increases with the $(\epsilon_e/\epsilon_B)$ ratio during both the fast and slow cooling regime, and that for a given$(\epsilon_e/\epsilon_B)$ value it is greater during the fast cooling regime than the slow cooling regime. We then evaluate IC flux as a function of $\epsilon_e$ and $\epsilon_B$ and we find that for typical GRB model parameters during the first phases of afterglow emission, e.g. $\sim$500 s, if the fireball is in the fast cooling regime, IC flux integrated over the full energy range can reach values as high as $\sim 10^{-8}~erg~ cm^{-2} s^{-1}$. However, at early times, the peak of the IC component can be at very high energies, well above the GeV range, depending on the position of the peak of the synchrotron component. We thus study the detectability of IC component as a function of $\epsilon_e$ and $\epsilon_B$ in the Large Area Telescope (LAT) above 1 GeV as a function of the integration time. We find that for a sufficiently long integration time, such as $\sim$10000 s, if the fireball is still in the fast cooling regime LAT can detect IC afterglow emission for a significantly larger region of parameters space. We develop a numerical model which permits to study ES synchrotron plus IC emission at times smaller than the fireball deceleration phase. We produce complete early-to-late time synchrotron and IC light curves and spectra, and compare them with the LAT sensitivity. As suggested by the studies performed in the $\epsilon_B$, $(\epsilon_e/\epsilon_B)$ parameters space, when most of the emission occurs during the fast cooling regime LAT can detect IC emission from the afterglow, and the detectability of this component increases with the $(\epsilon_e/\epsilon_B)$ ratio. One can easily extrapolate these results to the case of AGILE, where the effective area around 1 GeV is a factor of $\sim 15$ lower than LAT. The main conclusion is that we would expect detection by AGILE up to a redshift $z \sim$0.25; this is consistent with the findings of \citet{gou07}.

The large IC fluxes that can be produced in the afterglow phase motivated us to apply this model to the very bright \object{GRB 940217}, which presented high energy emission lasting from $\sim~ 500$ s to $\sim ~ 5000$ s after the burst. We find that IC emission from ES can account for this delayed emission. We note also that \citet{wei07} recently proposed an alternative scenario, in which this delayed high energy emission is ascribed to IC emission arising from a decelerating blast wave in the presence of a late energy injection plus temporal evolution of the shock parameters $\epsilon_e$ and $\epsilon_B$.

The case of DES (thick shell fireballs) produced by a long lasting central engine activity is interesting, because it offers a possible explanation for the origin of flares with a soft spectrum consistent with that of the subsequent afterglow emission, or that of flares showing a spectral evolution $\Delta \Gamma$ of the order of 0.5-1.0. In this context X-ray flares represent the onset of afterglow emission, thus the flare photons can be IC scattered by the afterglow electrons, producing flares in the GeV-TeV band. We calculate synchrotron and IC spectra and light curves, and find that bright GeV flares can be produced for a burst with fireball kinetic energy of the order of $\sim~10^{54}$ erg if the fireball is in the fast cooling regime at the time of the flare appearance. The analysis of a sub-sample of bursts with known redshift having flares in their X-ray light curve shows that $\sim$ 30 \% of them satisfy this requirement (see Table \ref{flare_properties}), and are expected to present detectable high energy flares. Short GRBs typically have lower isotropic energies, see e.g GRB050724 in Table \ref{flare_properties}, thus we do not expect detectable high energy flares from this population of bursts. A comparison of model predictions with the LAT sensitivity shows that LAT is able to detect high energy flares and IC emission arising from the afterglow up to $\sim$ 1 GeV if most of the emission occurs during the fast cooling regime, and that the detectability of the IC component increases with the $(\epsilon_e/\epsilon_B)$ ratio. Given an X-ray flare with mean fluence $\approx~3 \times 10^{-7}~erg~cm^{-2}$, if the Compton parameter $Y$ is of the order of unity, the expected high energy flare can be detected by LAT at 100 MeV, while its fluence is below the 1 GeV LAT threshold. However, we have shown that for typical model parameters $Y$ can be larger than unity, and in this case the fluence of the high energy flare overcomes the 1 GeV LAT threshold. We also estimate, under reasonable assumptions, the values of flux and peak energy of the expected high energy flare from those of the X-ray flare. This is a very important information because it allows us to understand whether an observed X-ray flares with certain properties can produce a detectable high energy flare.

In the DES scenario low and high energy flares are produced by the same electron population and source region, thus they are expected to be simultaneous once one considers their integrated broadband flux. This is an important prediction and represents one of the distinctive elements between the several models proposed to explain the origin of flares. If X-ray flares are produced by LIS associated with a long-lived central engine, then high energy flares can be produced through two different mechanisms. One possibility is that X-ray flare photons are IC scattered on the electrons accelerated by the forward shock (External Inverse Compton, EIC). In this case, due to the fireball curvature and the anisotropy of the incoming photons in the frame of forward shock electrons, the high energy flare is expected to extend up to longer time scales than the X-ray flare, and the flux received by an observer is lowered (see \citet{fan07} for details). The second possibility is that high energy flares are produced in LIS by SSC of the same electrons producing X-ray flares. In this case one expects a good temporal correlation between the X-ray and high energy flare, as in the framework of DES. However, as shown by \citet{fan07}, in the framework of LIS, SSC emission is expected to peak at lower energy with respect to the ES model. One should also note that to observe the simultaneity of low and high energy flares, the peak energies of both flares need to fall within or below the observational band of the instrument. We predict that X-ray flares with peak energy of $\sim$ 10 eV produce high energy flares with peak energy around hundred of MeV-GeV. Consequently, to distinguish between LIS and ES models requires both spectral and temporal information. Therefore future GLAST observations, and their coordination with those of the \emph{Swift} satellite, will play an important role in narrowing down models attempting to explain the origin of flares, and thus shedding light on the physics of the central engine.

\begin{acknowledgements}

The authors are grateful to F. Longo and N. Omodei for their support in the estimation of the LAT sensitivity and for helpful discussions. A.G also acknowledges B. Gendre for his support in the \emph{Swift} GRBs analysis.

\end{acknowledgements}


%

\end{document}